\documentclass[
twocolumn,
preprint, 
nofootinbib,
amsmath,amssymb,
aps, 
prd,
superscriptaddress,
floatfix,
]{revtex4-2}

\usepackage[T1]{fontenc}

\usepackage{graphicx}
\usepackage{dcolumn}
\usepackage{multirow}
\usepackage{makecell}
\usepackage{xurl}
\usepackage{amsmath}
\usepackage{longtable}
\usepackage{rotating}
\usepackage{bm}
\usepackage[colorlinks,
            linkcolor=red,
            anchorcolor=blue,
            citecolor=green
            ]{hyperref}
\usepackage{cleveref}
\crefformat{equation}{Eq.~(#2#1#3)}
\Crefformat{equation}{Eq.~(#2#1#3)}
\crefrangeformat{equation}{Eqs.~(#3#1#4)--(#5#2#6)}
\Crefrangeformat{equation}{Eqs.~(#3#1#4)--(#5#2#6)}
\crefformat{section}{Sect.~#2#1#3}
\Crefformat{section}{Sect.~#2#1#3}
\crefformat{figure}{Fig.~#2#1#3}
\Crefformat{figure}{Fig.~#2#1#3}
\crefformat{table}{Table~#2#1#3}
\Crefformat{table}{Table~#2#1#3}

\begin{document}


\title{\textbf{Bayesian Analysis of Gravitational Wave Microlensing Effects from Galactic Double White Dwarfs} 
}%

\author{Yan Sun}
\affiliation{School of Fundamental Phusics and Mathematical Sciences,
Hangzhou Institute for Advanced Study, University of Chinese Academy of Sciences (UCAS), Hangzhou 310124, China}
\affiliation{Center for Gravitational Wave Eaperiment, National Microgravity Laboratory, Institute of Mechanics, Chinese Academy of Sciences, Beijing 100190, China}
\affiliation{University of Chinese Academy of Sciences, Beijing 100049, China}

\author{Yong Yuan}
\email{yuanyong@imech.ac.cn}
\affiliation{Center for Gravitational Wave Experiment, National Microgravity Laboratory,
Institute of Mechanics, Chinese Academy of Sciences, Beijing 100190, China}

\author{Minghui Du}
\affiliation{Center for Gravitational Wave Experiment, National Microgravity Laboratory,
Institute of Mechanics, Chinese Academy of Sciences, Beijing 100190, China}

\author{Wen-Fan Feng}
\affiliation{Kavli Institute for Astronomy and Astrophysics,
Peking University, Beijing 100871, China}

\author{Xilong Fan}
\email{xilong.fan@whu.edu.cn}
\affiliation{School of Physics Science and Technology,
Wuhan University, Wuhan 430072, China}

\author{Peng Xu}
\email{xupeng@imech.ac.cn}
\affiliation{Center for Gravitational Wave Experiment, National Microgravity Laboratory,
Institute of Mechanics, Chinese Academy of Sciences, Beijing 100190, China}
\affiliation{Hangzhou Institute for Advanced Study, University of Chinese Academy of Sciences, Hangzhou 310124, China}
\affiliation{Taiji Laboratory for Gravitational Wave Universe (Beijing/Hangzhou),
University of Chinese Academy of Sciences, Beijing 100049, China}


\date{\today}
\begin{abstract}
Gravitational waves (GWs) from the galactic double white dwarf (DWD) systems are one of the primary targets for upcoming space-based detectors. Due to their vast abundance and widespread distribution throughout the Galactic disk and bulge, these systems may provide a high-statistical population for probing GW microlensing effects induced by Galactic compact objects. To evaluate the detectability of such effects, in this work we simulate the four-year observation of DWD systems by Taiji, in the form of a second-generation Time Delay Interferometry (TDI) data stream. Within a Bayesian inference framework, we estimate parameters for lensed GWs from DWD systems for different values of the lens parameters, including the lens mass $M_\mathrm{L}\in [10, 10^6]$\,M$_\odot$, the effective velocity $v_\mathrm{eff}\in [50, 500]$\,km/s and the initial separation $L\in [R_\mathrm{E}, 3R_\mathrm{E}]$, and obtain the uncertainties of the corresponding parameters. These results characterize the capability of future Taiji observations to probe such systems. We further employ the Bayesian model selection framework to distinguish between lensed and unlensed scenarios, and investigate the impacts of three key physical parameters of the lens system: $M_\mathrm{L}$, $v_\mathrm{eff}$, and $L$ on distinguishing lensing events. Our results show that when $M_\mathrm{L}$ is below $10^5$\,M$_\odot$ or $L\geq3R_\mathrm{E}$, it is not possible to distinguish between lensed and unlensed models. For $v_\mathrm{eff}$, although the Bayes factor decreases as $v_\mathrm{eff}$ decreases, the lensed and unlensed models can still be distinguished within our parameter range.
\end{abstract}

\maketitle

\begin{centering}
    \section{INTRODUCTION}
\end{centering}

Gravitational lensing was predicted by Einstein's theory of general relativity \cite{1916AnP...354..769E}. This effect  occurs when the trajectory of light is curved due to the gravity of a massive object located between the source and the observer. In the electromagnetic spectrum, gravitational lensing has been extensively observed and studied \cite{1979Natur.279..381W,1988Natur.333..537H,2011PhRvL.107b1301D,2005ApJ...625..633D,2024SSRv..220...87S}. Gravitational lensing can be produced by a wide range of cosmic structures, including galaxies \cite{2024SSRv..220...87S}, galaxy clusters \cite{2020Sci...369.1347M}, dark matter halos \cite{2010RSPTA.368..967E,2010PhRvD..82l3507W}, dark matter substructure \cite{2010AAS...21530003F,2013MNRAS.431.2172Z}, etc. The first gravitational lensing event was observed in 1979 \cite{1979Natur.279..381W}, the two objects 0957 + 561 A, B were regarded as a single source, later resolved into two images by gravitational lensing, directly confirming this prediction of general relativity. 

Similarly, extensive theoretical research on GW lensing has shown that GWs also display lensing effects analogous to those of light rays \cite{2020PhRvD.102l4048E,1996PhRvL..77.2875W,2010PhRvL.105y1101S,2018MNRAS.476.2220L,2018PhRvD..98j4029D}. These effects include the magnification of signals, the generation of multiple images, the introduction of relative time delays and distortions of the signal phase \cite{2003ApJ...595.1039T,1999PhRvD..59h3001B,1999PThPS.133..137N,1992grle.book.....S}. Collectively, these effects offer a unique probe for investigating the properties of the source and lensing object \cite{2025PhRvD.111h4067N,2026ApJ...997...11Y,2014PhRvD..90f2003C}. These theoretical works establish the methodological basis and feasibility for searches for gravitationally lensed GW events in data from current and future detectors.

The LIGO–Virgo–KAGRA (LVK) Collaboration  has already
detected a substantial number of GW events, originating from the coalescence of compact binary systems, such as binary black hole mergers, black hole–neutron star and binary neutron star mergers \cite{2016PhRvL.116x1102A, 2016PhRvL.116f1102A, 2016PhRvL.116x1103A, 2016PhRvL.116v1101A, 2017PhRvL.119p1101A, 2018PhRvL.121p1101A, 2023PhRvX..13d1039A, 2019ApJ...882L..24A, 2021PhRvX..11b1053A}. Although  current ground-based detectors are predicted to be capable of detecting GW events with lensing effects, none have been observed so far \cite{2018PhRvD..98h3005L}. Next-generation ground-based detectors, such as the Cosmic Explorer \cite{2017CQGra..34d4001A} and  the Einstein Telescope \cite{2010CQGra..27s4002P}, represent a significant improvement over the second-generation. According to theoretical estimates, the detection rate of GW events with lensing effects is expected to increase substantially \cite{2018MNRAS.476.2220L,2018PhRvD..98j4029D}.

Complementing these ground-based detectors, space-based GW detectors, such as LISA \cite{2017arXiv170200786A}, Taiji \cite{10.1093/nsr/nwx116} and TianQin \cite{2016CQGra..33c5010L}, are designed to detect GW in the milliHertz frequency band, effectively extending the detectable GW spectrum. To date, substantial work has already been done on the gravitational lensing effects anticipated for these space-based GW detectors \cite{2026ApJ...997...11Y,2025PhRvD.112l4019S,2023PhRvD.108l3543C,2025PhRvD.111b4068B} and the detection of lensed events using deep learning and machine learning \cite{2021PhRvD.104l4057G, 2021ApJ...915..119K, 2024MNRAS.535..990M}.

DWDs are one of the main target sources for space-based GW detectors \cite{2017arXiv170200786A,10.1093/nsr/nwx116,2016CQGra..33c5010L}, with an estimated $\sim10^4$ such systems expected to be observed (e.g.\cite{Du:2025xdq}). Within the Milky Way, a large population of moving astrophysical objects including black holes\cite{2024A&A...686L...2G,2023MNRAS.518.1057E,2020A&A...638A..94O,2024Natur.631..285H,2024ApJ...970...74P,2022MNRAS.516.4971S}, globular clusters\cite{1996AJ....112.1487H,2018MNRAS.478.1520B,2023ApJ...956...70P,2024A&A...687A.214G,2010arXiv1012.3224H,2005ApJS..161..304M}, and dark matter subhalos\cite{Springel:2008cc,2007ApJ...667..859D,2008Natur.454..735D} can act as gravitational lenses and potentially induce microlensing (small lens size compared to the GW wavelength) \cite{2023MNRAS.526.3832J} effects on DWDs signals. The relative motion between these lenses and the DWD sources induces a time-dependent lensing geometry, which imprints characteristic diffraction or
interference fringes onto the detected GW signals \cite{2019ApJ...875..139L}. Detecting such lensing events could provide a direct probe of black hole formation \cite{2022ApJ...933...83S,2022ApJ...932L...4G}. In addition, lensing by dark matter subhalos would probe the small-scale structure of dark matter in the Milky Way and test the cold dark matter paradigm \cite{Springel:2008cc,1999ApJ...524L..19M,1996ApJ...462..563N}. Motivated by such a prospect, in this work we focus on the detection capability of the future Taiji observatory for these effects.

In this paper, we study gravitationally lensed GW signals from DWD, performing parameter estimation and Bayesian model selection between lensed and unlensed hypotheses.
We model the potential lenses as point mass lenses (PML), a framework that effectively characterizes a diverse population of Galactic objects. To improve computational efficiency and facilitate Bayesian inference, we employ an accelerated method to calculate DWD waveforms based on the principle of fast-slow decomposition and further extend to second-generation TDl and lensed systems \cite{2007PhRvD..76h3006C,2025PhRvD.112j2007C}. Subsequently, we explore the parameter space for the lens mass $M_\mathrm{L}\in [10, 10^6]$\,M$_\odot$, the relative transverse velocity $v_\mathrm{eff}\in [50, 500]$\,km/s of the source with respect to the lens, and the initial position $L\in [R_\mathrm{E}, 3R_\mathrm{E}]$ relative to the closest point to the lens, in order to investigate the selection of the model between the lensed and unlensed hypotheses across this parameter space. This velocity range $v_\mathrm{eff}\in [50, 500]$\,km/s is based on the dynamics of the galaxy \cite{2025PhRvD.112l4019S,1988gady.book.....B}.
Throughout this study, we adopt natural units where $c=G=1$.

This paper is organized as follows. In \Cref{sec:model}, we introduce the wave optic formalism for GW lensing and the Bayesian framework adopted in this work. \Cref{sec:Data Analysis} describes the simulated data sets, as well as the parameter estimation and model comparison results for different lens configurations. Finally, we summarize our conclusions in \Cref{Sec:Conclusion}.

\begin{centering}
    \section{MODEL And Method}
    \label{sec:model}
\end{centering}

\begin{centering}
    \subsection{Gravitational Lensing}
\end{centering}

\begin{figure}[htbp]
    \centering
    \includegraphics[width=\linewidth]{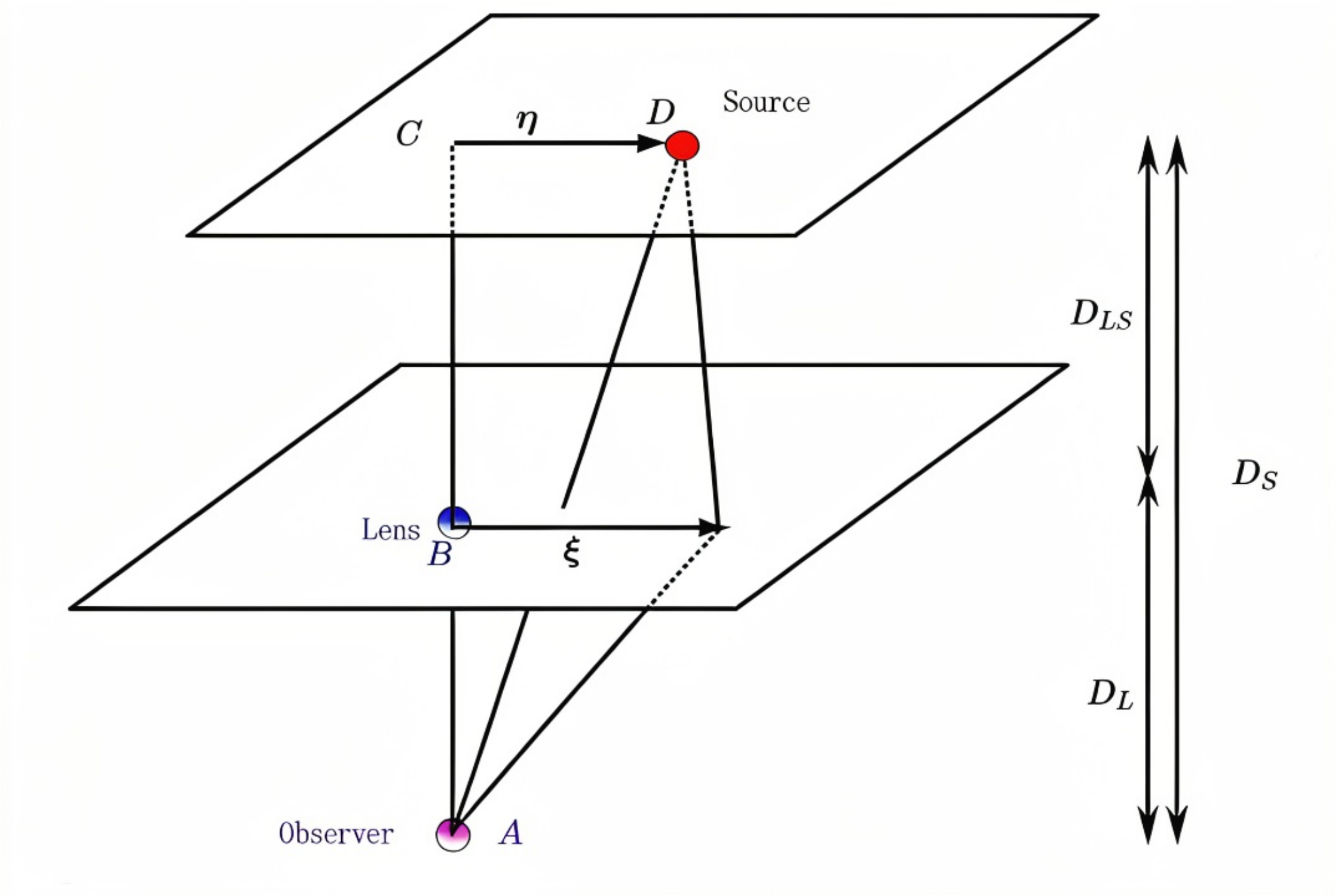}
    \caption{\footnotesize Schematic illustration of a gravitational lensing system. The source plane (top) emits GW that is deflected by the intervening lensing mass $M_\mathrm{L}$ at the lens plane (center) and the observer at the bottom. }
    \label{fig:1}
\end{figure}

In this section, we will introduce the base concepts of gravitational lensing based on the works in Refs \cite{2003ApJ...595.1039T,2019ApJ...875..139L}. The geometry of a moving lensing system is illustrated in \Cref{fig:1}. Where $D_\mathrm{L}$, $D_\mathrm{S}$, $D_\mathrm{LS}$ denote the angular diameter distances to the lens,  to the source and from the source to the lens, respectively, and $\bm{\eta}$ is the position vector of the source in the source plane, $\bm{\xi}$ is the impact parameter in the lens plane. 

The effect of lensing on a GW is defined as the ratio of its lensed amplitude to its unlensed amplitude \cite{2003ApJ...595.1039T, 1992grle.book.....S, 2025PhRvD.112l4019S}. This ratio is known as the amplification factor and is given by:
\begin{equation}
    F(f, \boldsymbol{y}) = \frac{D_\mathrm{S}\xi_0^2}{D_\mathrm{L} D_\mathrm{LS}} \frac{f}{i} \int d^2\boldsymbol{x} \, \exp[2\pi i f t_\mathrm{d}(\boldsymbol{x},\boldsymbol{y})].
\end{equation}
Here, $\xi_0=R_\mathrm{E}$ is the Einstein radius of the lens, and $\boldsymbol{x}=\boldsymbol{\xi}/\xi_0,\boldsymbol{y}=\boldsymbol{\eta}D_\mathrm{L}/\xi_0D_\mathrm{S}$ is the dimensionless vectors normalized by $\xi_0$. And $f$ represents the source's GW frequency, $t_\mathrm{d}$ denotes the time delay:
{\small
\begin{equation}
t_\mathrm{d}(\boldsymbol{x},\boldsymbol{y})=\frac{D_\mathrm{S}\xi_0^2}{D_\mathrm{L}D_\mathrm{LS}}\left[\frac{1}{2}\left|\boldsymbol{x}-\boldsymbol{y}\right|^2-\psi(\boldsymbol{x})+\phi_\mathrm{m}(\boldsymbol{y})\right],
\end{equation}
}
where $\psi(\boldsymbol{x})$ is the deflection potential, $\phi_\mathrm{m}(\boldsymbol{y})$ is chosen so that  the minimum value of the arrival time is zero. 

Our study models the lensing objects as PML to simplify the analysis. We use \textsc{GLoW} module \cite{2025PhRvD.111j3539V} to compute the amplification factor $|F|$ and its phase $\theta_\mathrm{F}$. The corresponding density of surface mass is $\Sigma(\xi)=M_\mathrm{L}\delta^2(\xi)$, and the lensing potential is $\psi(\boldsymbol{x})=\ln x$, $\phi_\mathrm{m}(y)=\left(x_\mathrm{m}-y\right)^2/2-\ln x_\mathrm{m}$ with $x_\mathrm{m}=[y+(y^2+4)^{1/2}]/2$, where $x\equiv|\boldsymbol{x}|$ and $y\equiv|\boldsymbol{y}|$. In this case, the amplification factor is as follows: 
\begin{equation}
\begin{aligned}
F(w, y)&=\exp\left\{\frac{\pi w}{4}+i\frac{w}{2}\left[\ln\left(\frac{w}{2}\right)-2\phi_\mathrm{m}(y)\right]\right\}\\&\times\Gamma\left(1-\frac{i}{2}w\right) {}_1F_1\left(\frac{i}{2}w,1;\frac{i}{2}wy^2\right),
\end{aligned}
\end{equation}
where $w=8\pi M_\mathrm{L}f$, $_{1}F_{1}$ is the confluent hypergeometric function, $\Gamma$ is the Gamma function. The Einstein radius of the PML is the following:
\begin{equation}
\xi_0\equiv\left(\frac{4M_\mathrm{L}D_\mathrm{L}D_\mathrm{LS}}{D_\mathrm{S}}\right)^{1/2}.
\end{equation}

\begin{figure}[htbp]
    \centering
    \includegraphics[width=\linewidth]{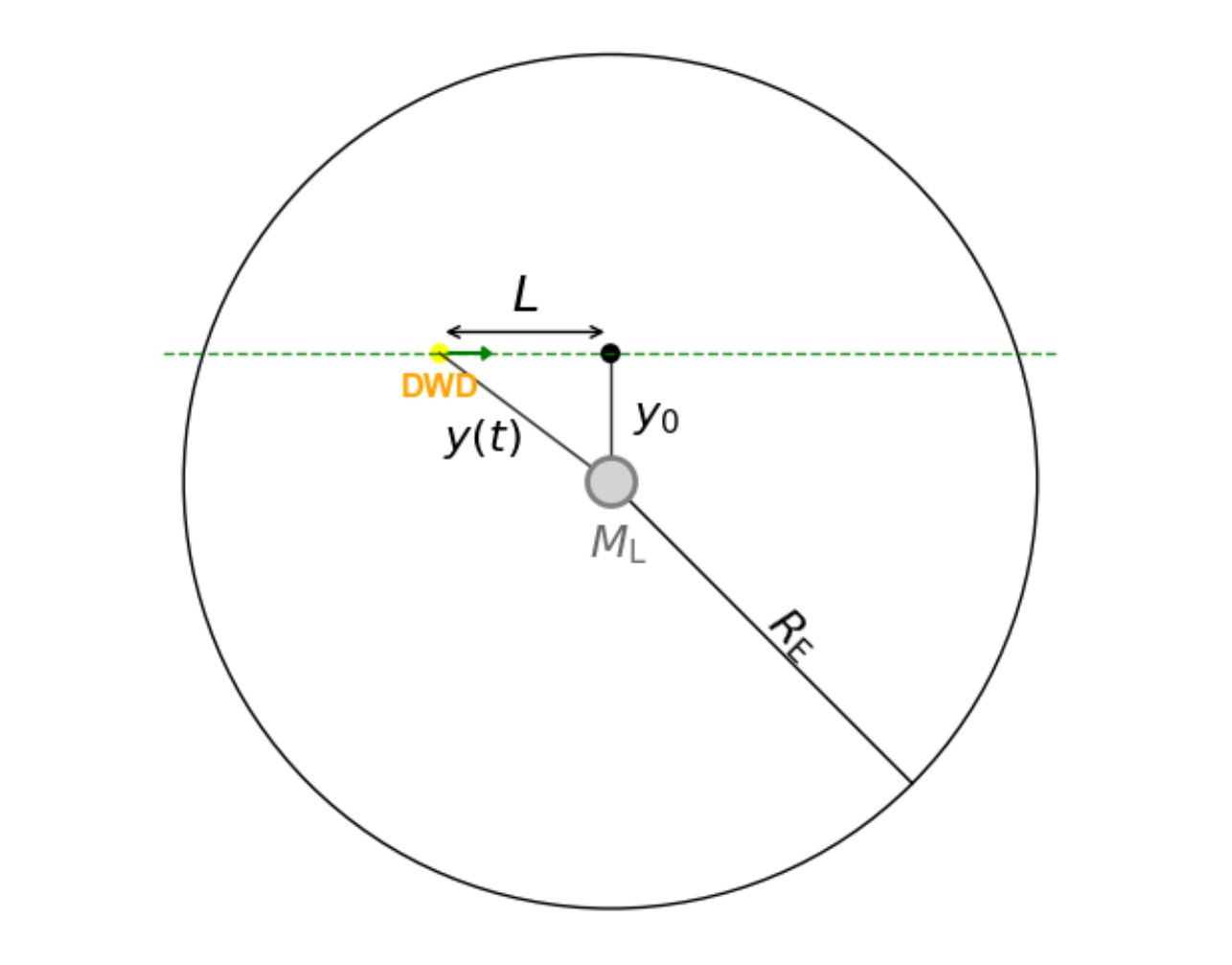}
    \caption{\small Schematic diagram illustrating a moving source and a stationary lens.}
    \label{fig:2}
\end{figure}

In this work, we consider the relative motion of the source with respect to the lens, as shown in \Cref{fig:1} and \Cref{fig:2}. Here, $v_\mathrm{eff}$ denotes the relative velocity between the source and the lens, and $y_0$ represents the closest distance between the lensing object and the source at time $t_0$. In this case, the impact parameter $y$ is a function of time and can be written as: \cite{2019ApJ...875..139L}:
\begin{equation}
y(t)=\sqrt{y_0^2+\left(\frac{t-t_0}{t_\mathrm{E}}\right)^2},
\label{eq.5}
\end{equation}
where $t_\mathrm{E}=\frac{R_\mathrm{E}}{v_\mathrm{eff}}$ is the  Einstein crossing time. We rewrite \Cref{eq.5} as follows:
\begin{equation}
y(t; v_\mathrm{eff}, L)=\sqrt{y_0^2+\left(\frac{t\cdot v_\mathrm{eff}-L}{R_\mathrm{E}}\right)^2},
\end{equation}
where $L$ denotes the initial position corresponding to the beginning of the observation period as shown in \Cref{fig:2} . $T$ denotes the total observation time and is set to four years in this work, and $y_0 = 0.1$. Accordingly, the amplification factor can be written as:
\begin{equation}
\begin{aligned}
&F(f,t;M_\mathrm{L},L,v_\mathrm{eff})\\=&\exp\left\{\frac{\pi w}{4}+i\frac{w}{2}\left[\ln\left(\frac{w}{2}\right)-2\phi_\mathrm{m}(y(t))\right]\right\}\\&\times\Gamma\left(1-\frac{i}{2}w\right) {}_1F_1\left(\frac{i}{2}w,1;\frac{i}{2}wy^2(t)\right).
\label{Eq.7}
\end{aligned}
\end{equation}

\begin{centering}
    \subsection{Bayesian Analysis}
\end{centering}

In this section, we describe the Bayesian inference framework.
Our Bayesian analysis is performed exclusively using orthogonal noise TDI channels $A_\mathrm{2}$ and $E_\mathrm{2}$. Setting the total observation duration to four years with a sampling frequency of 0.1 is sufficient and ensures computational efficiency.
Simulated GW signals without lensing in the frequency domain can be expressed as follows:
\begin{equation}d(f;\theta^U)=h(f;\theta^U)+n(f),\end{equation}
where $h(f;\theta^U)$ represents the unlensed GW signal, $n(f)$ denotes the Taiji detector noise,  and $\theta^U$ denotes the vector of source parameters.

For GW signals that can be modeled as monochromatic, the modulation effect in \Cref{Eq.7} is effectively characterized by the time-dependent factor $F(t;M_\mathrm{L},L,v_\mathrm{eff})$, so that the simulated lensed GW signals under a PML in the frequency domain are given by:
\begin{equation}
\begin{aligned}
d^L(f;\theta^L,\theta^U)&=F(f;\theta^L)h(f;\theta^U)+n(f)\\&=h^L(f;\theta^L,\theta^U)+n(f),
\label{eq.9}
\end{aligned}
\end{equation}
where $F(f;\theta^L)$ is the amplification factor and $\theta^L = (M_\mathrm{L}, v_\mathrm{eff})$ denotes the lens parameter vector.

Using simulated lensed and unlensed GW data, we employ a Bayesian framework to systematically estimate parameters for both lensed and unlensed GW events. The Bayesian formulations for the lensed and unlensed cases are expressed as follows:
\begin{equation}
\begin{aligned}
    &p(\theta^U,\theta^L|d^L,M^L)\\=&\frac{\mathcal{L}(d^L|\theta^U,\theta^L,M^L)p(\theta^U,\theta^L|M^L)}{\mathcal{Z}^L(d^L|M^L)},
\end{aligned}
\end{equation}
\begin{equation}p(\theta^U|d^L,M^U)=\frac{\mathcal{L}(d^L|\theta^U,M^U)p(\theta^U|M^U)}{\mathcal{Z}^U(d^L|M^U)},\end{equation}
where $M^L, M^U$ denote the lensed and unlensed models, respectively. $\mathcal{L}$ is the likelihood function, $p(\theta|M)$ represents the prior distribution, $p(\theta|d,M)$ posterior distributions, and $\mathcal{Z}^L, \mathcal{Z}^U$ denotes Bayesian evidence for lensed and unlensed models, respectively. The evidence values for the lensed and unlensed models are given by:
\begin{equation}
\begin{aligned}
    \mathcal{Z}^L&=\int\mathcal{L}(d^L|\theta^U,\theta^L,M^L)\\&\times p(\theta^U,\theta^L|M^L)d\theta^Ud\theta^L,
\end{aligned}
\end{equation}
\begin{equation}\mathcal{Z}^U=\int\mathcal{L}(d^L|\theta^U,M^N)p(\theta^U|M^U)d\theta^U.\end{equation}

To compare the two hypotheses, we infer the parameters of the lensed data under both the lensed and unlensed models. The two hypotheses are assigned equal prior probabilities, with the additional constraint that any shared parameters are given identical prior distributions. Then, we use the Bayesian evidence of the two models to compute the Bayes factor, which evaluates the relative support for the hypothesis that the observed data are generated by a lensed versus an unlensed source, and can be expressed as:
\begin{equation}\ln\mathcal{B}^{L/U}=\ln\mathcal{Z}^L-\ln\mathcal{Z}^U.\end{equation}

\begin{centering}
    \section{Data Analysis}
    \label{sec:Data Analysis}
\end{centering}

\begin{centering}
    \subsection{Data Simulation And Analysis}
    \label{sec:III A.}
\end{centering}

\begin{figure}[htbp]
    \centering
    \includegraphics[width=\linewidth]{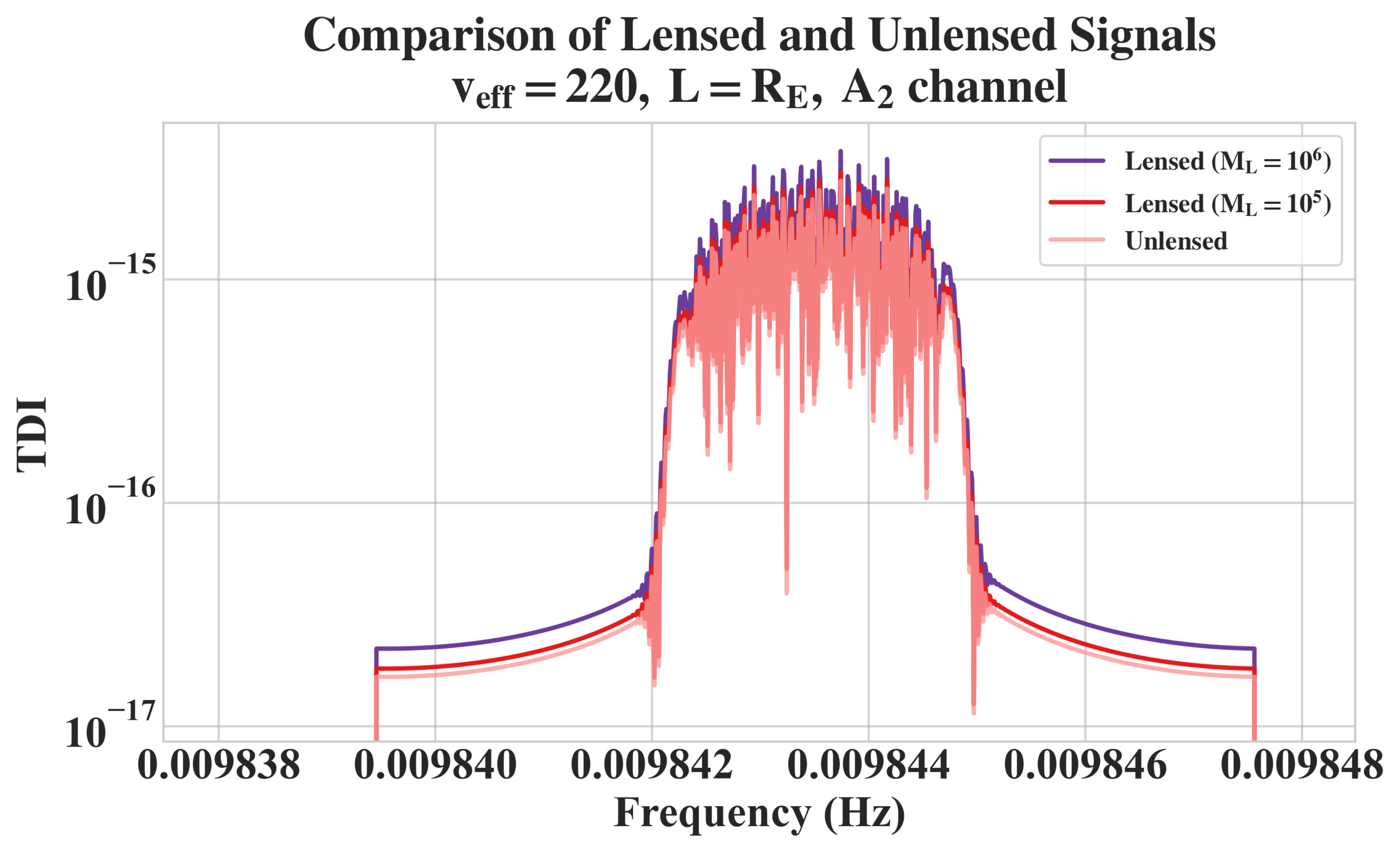}
    \includegraphics[width=\linewidth]{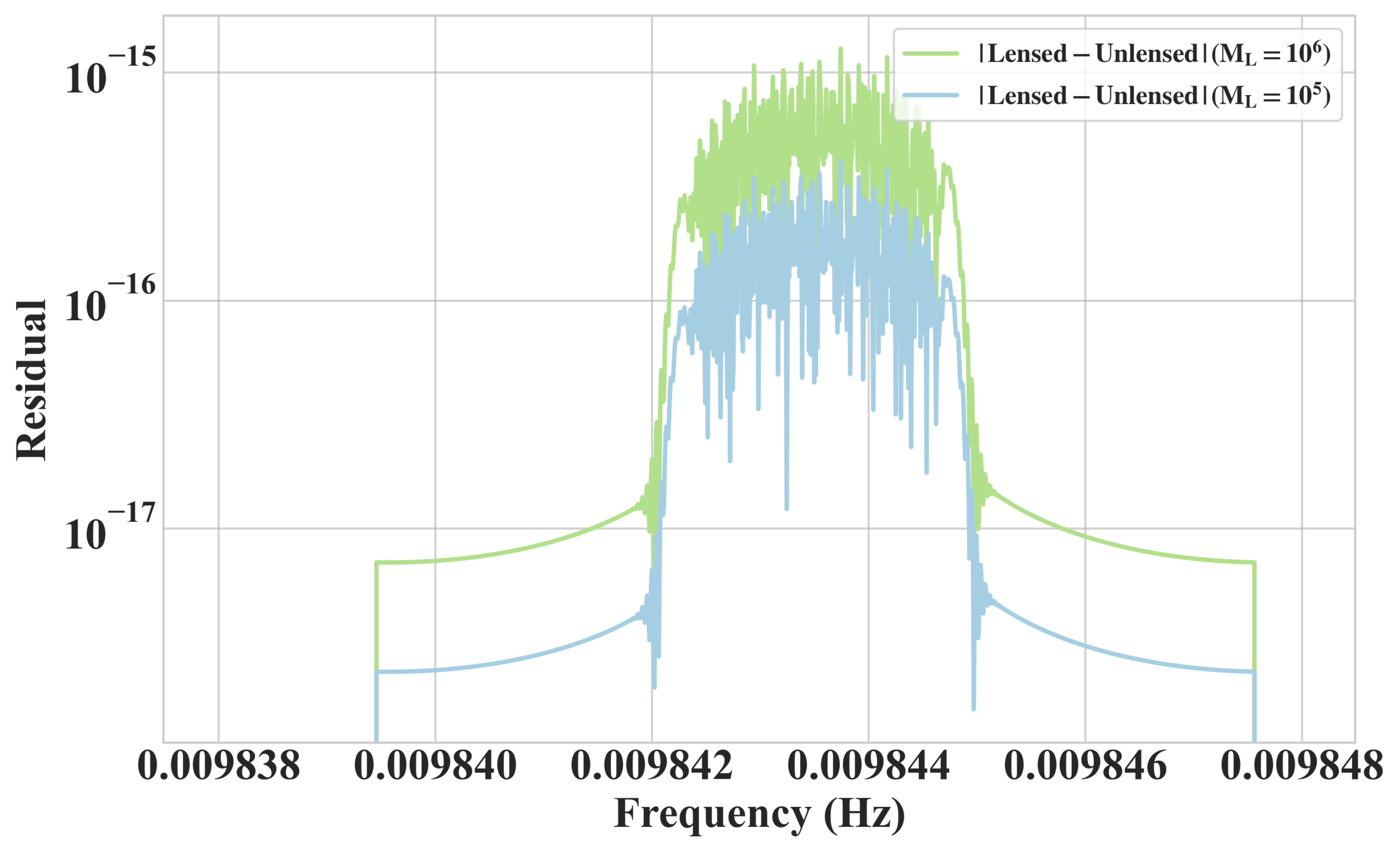}
    \caption{\footnotesize Comparison of lensed and unlensed signals for different $M_\mathrm{L}$. The top panel shows the comparison between the unlensed signal and the lensed signals at different $M_\mathrm{L}$. The bottom panel shows the residuals between the lensed and unlensed signals for different $M_\mathrm{L}$.}
    \label{fig:3}
\end{figure}

Within the PML model, we simulate lensed GW signals from DWDs according to \Cref{eq.9}. The GW singals form DWD can be written as \cite{2007PhRvD..76h3006C,2025PhRvD.112j2007C}:
\begin{equation}
\begin{aligned}
h_+(t)&=\mathcal{A}(1+\cos^2\iota)\cos\varphi\left(t\right),\\h_\times(t)&=2\mathcal{A}\cos\iota\sin\varphi\left(t\right),
\label{Eq15}
\end{aligned}
\end{equation}
\begin{equation}
\varphi(t)=2\pi\left(f_0t+\frac{1}{2}\dot{f}_0t^2+\frac{1}{6}\ddot{f}_0t^3\right)+\varphi_0.
\label{Eq16}
\end{equation}
We rewrite the polarizations in a more concise form for convenience of deduction (see Appendix~\ref{appA}): 
\begin{equation}
h_P(t)=\Re\left\{\mathcal{A}_Pe^{i\varphi(t)}\right\},\quad P\in\{+,\times\}
\label{Eq17}
\end{equation}
where
\begin{equation}
    \mathcal{A}_+\equiv \mathcal{A}(1+\cos^2\iota),\quad \mathcal{A}_\times \equiv-2i\mathcal{A}\cos\iota.
    \label{Eq18}
\end{equation}

We begin by examining the lensed and unlensed waveforms to analyze their differences. 
Then, we use \Cref{eq.9} to generate lensed signals within the TDI framework. Details of the accelerated waveform computation and the associated TDI calculations are presented in Appendix~\ref{appA}. For clarity, \Cref{fig:3} uses the $A_\mathrm{2}$ channel waveform as a representative case to show illustrative examples of the characteristic strain comparison between lensed and unlensed signals. The injection parameters for all examples shown in \Cref{fig:3} are performed assuming $(\mathcal{A}, f_0, \dot f_0, \iota, \varphi_0, \lambda, \beta, \psi)$ = ($1.40\times10^{-22}$, $9.84\times10^{-3}$, $8.31\times 10^{-15}$, $2.30$, $0.8$, $4.51$, $-0.3$, $1.38$). Here, $\mathcal{A}$ is the amplitude, $f_0$ is the frequency, $\dot f_0$ is the frequency derivative, $\iota$ is the inclination angle of the binary, $\varphi_0$ is the initial phase, $\lambda$ and $\beta$ are the ecliptic longitude and latitude of the source in the sky, and $\psi$ is the polarization angle. In principle, the luminosity distance from the source to the observer can be determined by $\mathcal{A}$, $f_0$, and $\dot f_0$. Since our research is confined to the Milky Way, where the cosmological redshift is negligible, the luminosity distance is treated as equivalent to the angular diameter distance. We set the lens plane to be located midway between the source plane and the observer plane in all the cases.
\begin{figure}[htbp]
    \centering
    \includegraphics[width=\linewidth]{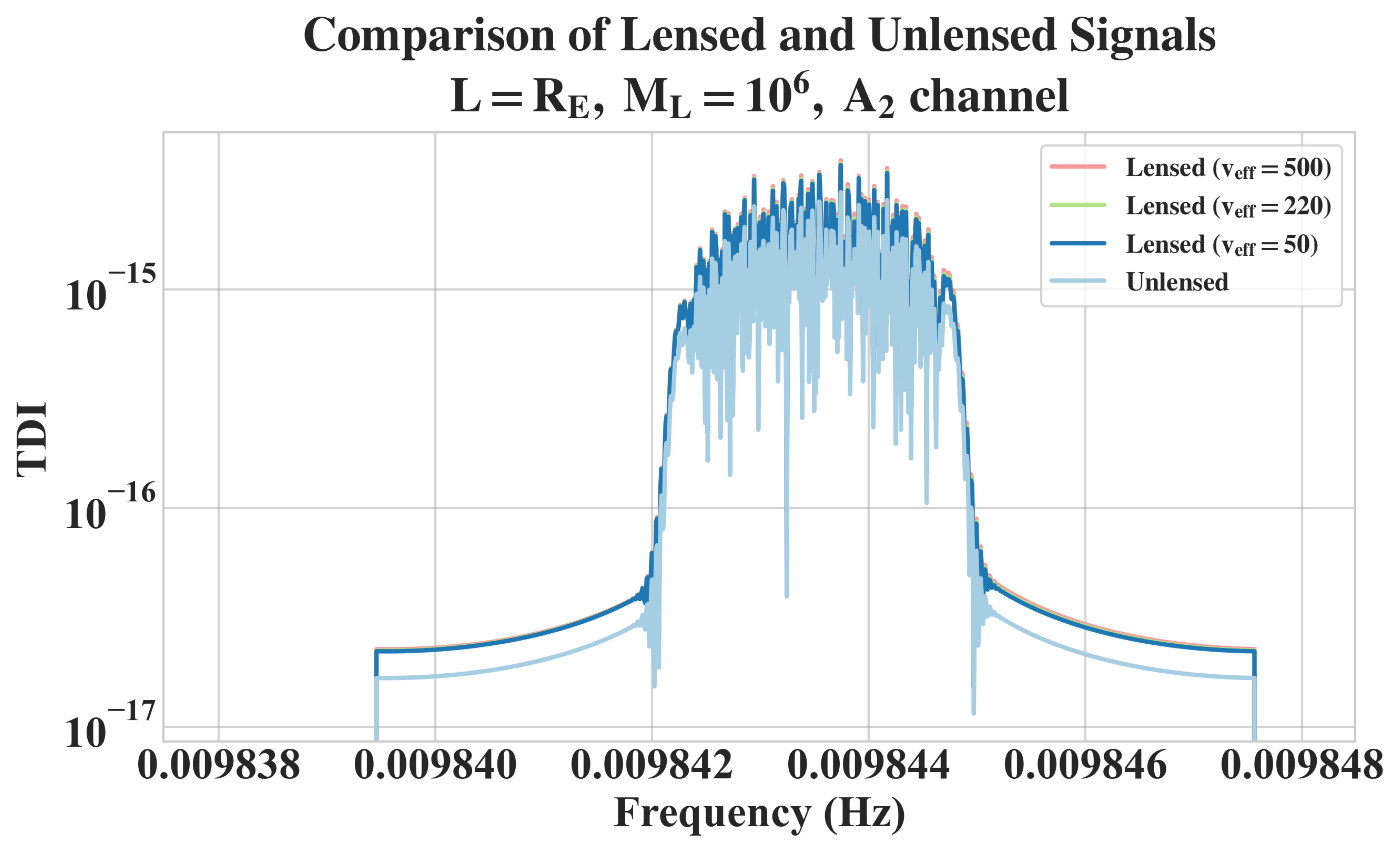}
    \includegraphics[width=\linewidth]{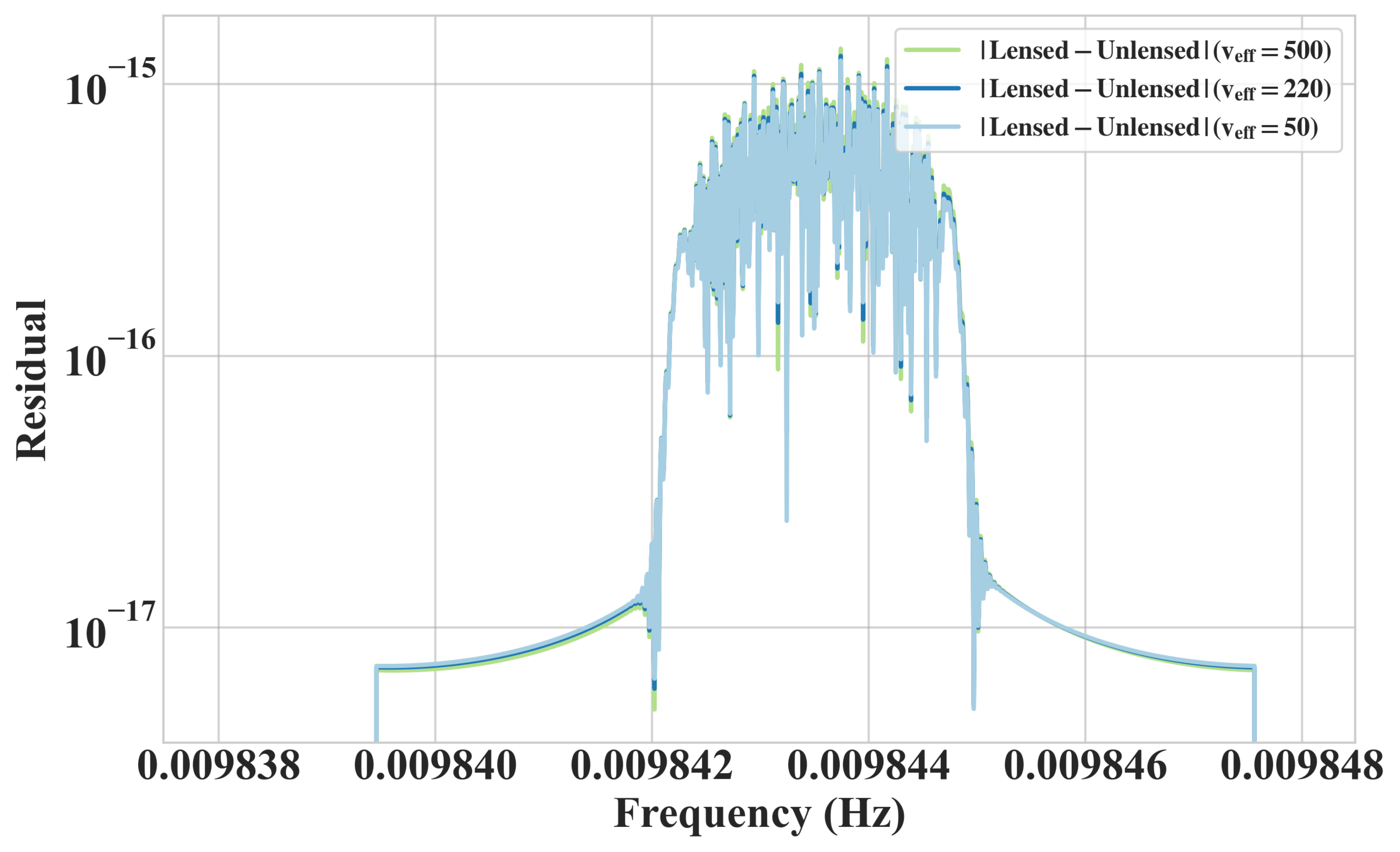}
    \caption{\footnotesize Effects of varying $v_\mathrm{eff}$ on the lensed signal. The top panel shows the unlensed signals and the lensed signals at different $v_\mathrm{eff}$. The bottom panel shows residuals between the lensed signals at different $v_\mathrm{eff}$ and the unlensed signal.}
    \label{fig:4}
\end{figure}

To investigate the impact of $M_\mathrm{L}$, we fixed $v_\mathrm{eff} = 220$\, km/s, $L = R_\mathrm{E}$, and varied $M_\mathrm{L}$ between $10^5$\,M$_\odot$ and $10^6$\, M$_\odot$. 
The upper panel of \Cref{fig:3} shows the resulting lensed and unlensed waveforms in the frequency domain, where stronger modulation near the central spectral peak is observed for larger $M_\mathrm{L}$. To quantify these differences, the bottom panel of \Cref{fig:3} presents the residuals between the lensed and unlensed signals for different $M_\mathrm{L}$. As expected, the amplitude of the residuals increases with increasing $M_\mathrm{L}$, confirming that larger lenses induce a stronger impact on the observed waveform.

The top panel of \Cref{fig:4} explores the effects of varying $v_\mathrm{eff}$ in the range $50,\, 220$ and $500$\,km/s on the lensed signals, while maintaining $M_\mathrm{L} = 10^6$\,M$_\odot$ and $L = R_\mathrm{E}$. As shown, the lensed signals corresponding to different effective velocities almost completely overlap in the frequency domain, indicating that variations in $v_\mathrm{eff}$ result in only slight differences in the overall waveform under this configuration. The bottom panel shows that the residuals between the lensed signals at different $v_\mathrm{eff}$ and the unlensed waveform are nearly the same, further validating this result. 
\begin{figure}[htbp]
    \centering
    \includegraphics[width=\linewidth]{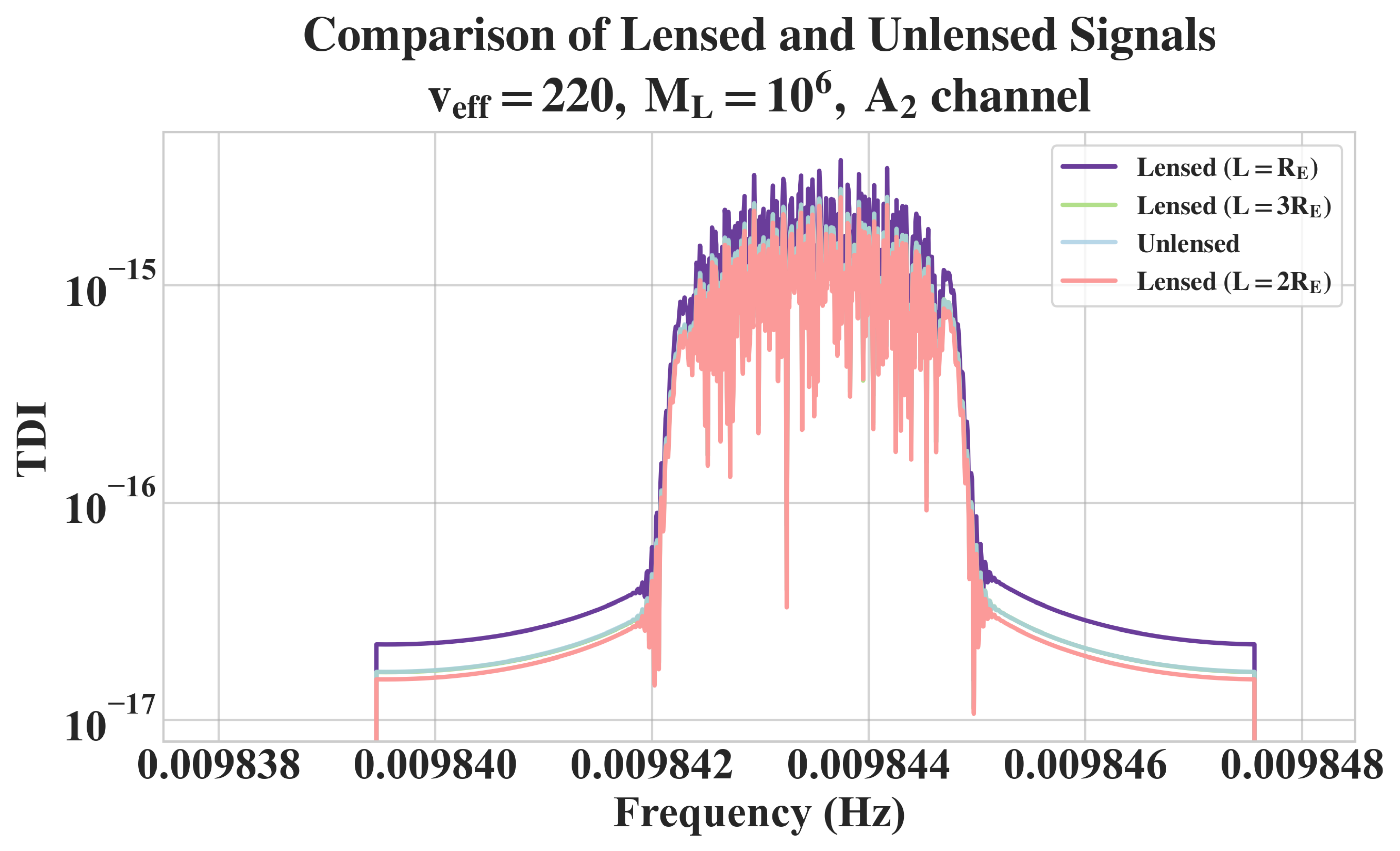}
    \includegraphics[width=\linewidth]{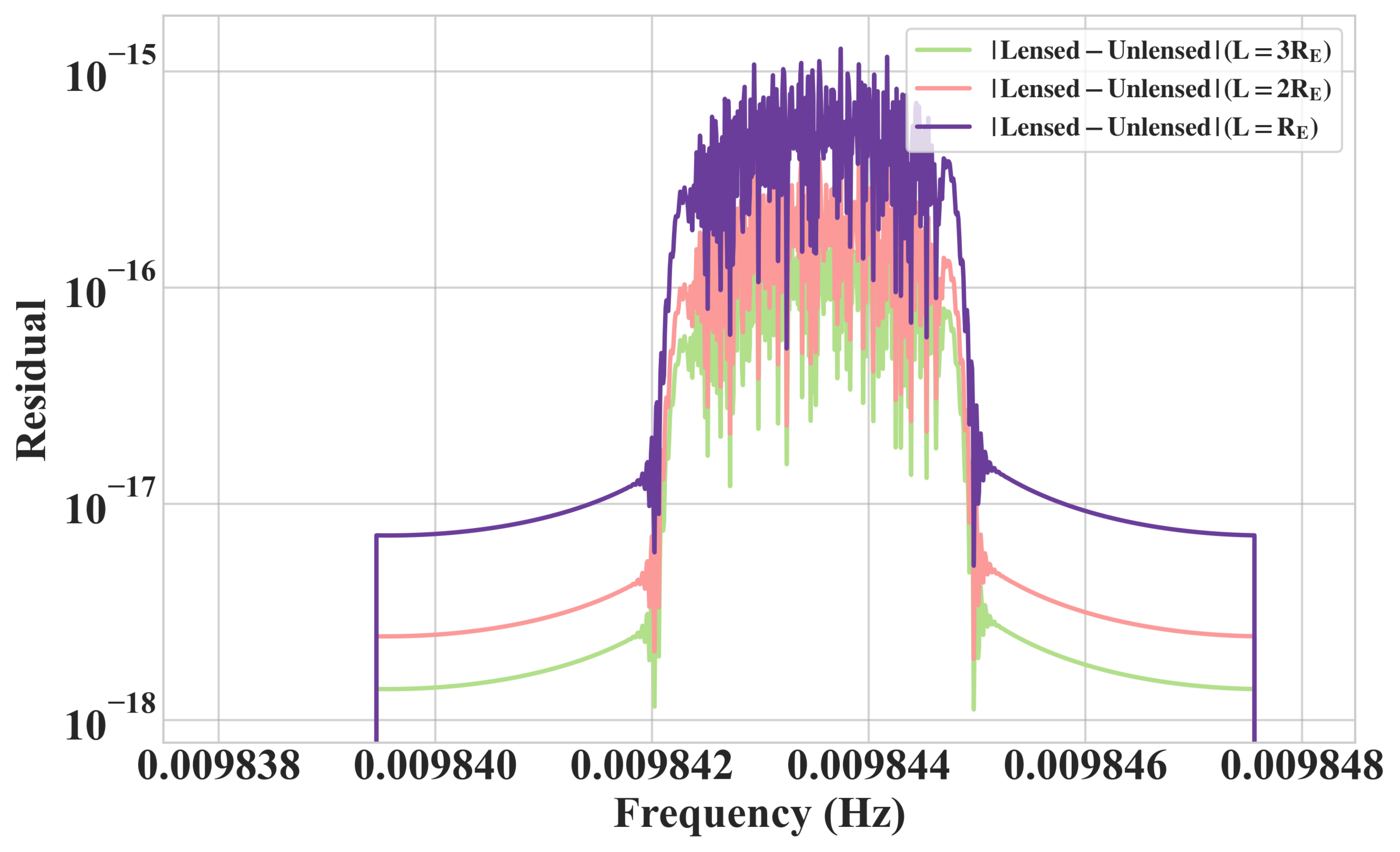}
    \caption{\footnotesize Signal variations for different initial positions $L = R_\mathrm{E},\, 2R_\mathrm{E},\, 3R_\mathrm{E}$. The top panel shows the unlensed signal and the lensed signal variations for each initial position $L$, while the bottom panel presents the residuals between the lensed and unlensed signals.}
    \label{fig:5}
\end{figure}

The signal variations induced by different initial positions $L =R_\mathrm{E},\, 2R_\mathrm{E},\, 3R_\mathrm{E}$ are shown in the top panel of \Cref{fig:5}, for a fixed $M_\mathrm{L} = 10^6$\,M$_\odot$ and $v_\mathrm{eff}=220$\,km/s. As illustrated, a decrease in $L$ corresponds to a smaller impact parameter $y$, thereby enhancing the lensing effects. To quantify these differences, the bottom panel presents the residuals between each lensed signal and the unlensed waveform. The results show that the amplitude of the residuals increases systematically from $L = 3R_\mathrm{E}$ to $L = R_\mathrm{E}$.

These analyzes demonstrate that variations in $M_\mathrm{L}$ and $L$ have a significant impact on the lensed waveforms,while $v_\mathrm{eff}$ plays a minimal role in shaping them.
\begin{sidewaystable*}[p]
\vspace{200pt}
\centering
\setlength{\abovecaptionskip}{20pt}
\setlength{\belowcaptionskip}{10pt}
\caption{Parameter estimation results for lensing cases.}
\label{tab:1}
\footnotesize
\begin{tabular}{c c c c c c c c c c c c c c}
\hline
Case & Model & $\mathcal{A}$ & $f_0$ & $\dot f_0$ & $\iota$ & $\varphi_0$ & $\lambda$ & $\beta$ & $\psi$ & $M_\mathrm{L}$ (M$_\odot$) & $v_\mathrm{eff}$ (km/s) & $\ln\mathcal{B}^{L/U}$ & SNR \\
\hline
\multirow{2}{*}{\makecell[c]{%
1: $M_\mathrm{L}=10^5$,\\
$v_{\rm eff}=220$, $L=R_\mathrm{E}$}}
 & Lensed & \begin{tabular}{@{}c@{}}$1.53 \times 10^{-22}$\\$\,^{+1.05 \times 10^{-24}}_{-1.15 \times 10^{-24}}$\end{tabular} 
& \begin{tabular}{@{}c@{}}$9.84 \times 10^{-3}$\\$\,^{+2.72 \times 10^{-11}}_{-2.69 \times 10^{-11}}$\end{tabular}
& \begin{tabular}{@{}c@{}}$8.31 \times 10^{-15}$\\$\,^{+4.02 \times 10^{-19}}_{-4.16 \times 10^{-19}}$\end{tabular}
 & \begin{tabular}{@{}c@{}}$2.29$\\$\,^{+6.55 \times 10^{-3}}_{-6.54 \times 10^{-3}}$\end{tabular}
&\begin{tabular}{@{}c@{}}$3.81 \times 10^{0}$\\$\,^{+0.0450}_{-3.12}$\end{tabular}
& \begin{tabular}{@{}c@{}}$4.51$\\$\,^{+9.18 \times 10^{-5}}_{-9.09 \times 10^{-5}}$\end{tabular}
& \begin{tabular}{@{}c@{}}$-0.301$\\$\,^{+2.11 \times 10^{-4}}_{-2.05 \times 10^{-4}}$\end{tabular}
 & \begin{tabular}{@{}c@{}}$2.94$\\$\,^{+0.197}_{-1.56}$\end{tabular} 
 & \begin{tabular}{@{}c@{}}$370$\\$\,^{+4.56 \times 10^3}_{ -3.38 \times 10^2}$\end{tabular}
 & \begin{tabular}{@{}c@{}}$298$\\$\,^{+196}_{-197}$\end{tabular}
 & 0.65 & 652 \\
  & Unlensed & \begin{tabular}{@{}c@{}}$1.54 \times 10^{-22}$\\$\,^{+9.55 \times 10^{-25}}_{-9.69 \times 10^{-25}}$\end{tabular} 
  & \begin{tabular}{@{}c@{}}$0.00984$\\$\,^{+2.62 \times 10^{-11}}_{-2.77 \times 10^{-11}}$\end{tabular} 
  & \begin{tabular}{@{}c@{}}$8.31 \times 10^{-15}$\\$\,^{+4.22 \times 10^{-19}}_{-4.01 \times 10^{-19}}$\end{tabular} 
  & \begin{tabular}{@{}c@{}}$2.29$\\$\,^{+6.81 \times 10^{-3}}_{-6.48 \times 10^{-3}}$\end{tabular}
  & \begin{tabular}{@{}c@{}}$3.80$\\$\,^{+0.0475}_{-3.11}$\end{tabular}
  &\begin{tabular}{@{}c@{}}$4.51$\\$\,^{+8.88 \times 10^{-5}}_{-9.37 \times 10^{-5}}$\end{tabular} 
  & \begin{tabular}{@{}c@{}}$-0.301$\\$\,^{+2.07 \times 10^{-4}}_{-2.10 \times 10^{-4}}$\end{tabular} 
  & \begin{tabular}{@{}c@{}}$2.93$\\$\,^{+0.0226}_{-1.56}$\end{tabular} & -- & -- & -- & -- \\
\hline
\multirow{2}{*}{\makecell[c]{%
2: $M_\mathrm{L}=10^6$,\\
$v_{\rm eff}=220$, $L=R_\mathrm{E}$}} & Lensed & \begin{tabular}{@{}c@{}}$1.41 \times 10^{-22}$\\$\,^{+4.14 \times 10^{-23}}_{-8.65 \times 10^{-24}}$\end{tabular} & 
\begin{tabular}{@{}c@{}}$0.00984$\\$\,^{+3.35 \times 10^{-11}}_{-3.85 \times 10^{-11}}$\end{tabular} & 
\begin{tabular}{@{}c@{}}$8.31 \times 10^{-15}$\\$\,^{+3.44 \times 10^{-19}}_{-3.28 \times 10^{-19}}$\end{tabular} & 
\begin{tabular}{@{}c@{}}$2.30$\\$\,^{+5.60 \times 10^{-3}}_{-5.39 \times 10^{-3}}$\end{tabular} & 
\begin{tabular}{@{}c@{}}$0.977$\\$\,^{+3.02}_{-0.278}$\end{tabular} & 
\begin{tabular}{@{}c@{}}$4.51$\\$\,^{+7.17 \times 10^{-5}}_{-7.24 \times 10^{-5}}$\end{tabular} & 
\begin{tabular}{@{}c@{}}$-0.301$\\$\,^{+1.75 \times 10^{-4}}_{-1.63 \times 10^{-4}}$\end{tabular} & 
\begin{tabular}{@{}c@{}}$1.40$\\$\,^{+1.56}_{-0.0174}$\end{tabular} & 
\begin{tabular}{@{}c@{}}$1020000$\\$\,^{+3.63 \times 10^{5}}_{-2.46 \times 10^{5}}$\end{tabular} & 
\begin{tabular}{@{}c@{}}$206$\\$\,^{+144}_{-93.8}$\end{tabular} & 43 & 804 \\
  & Unlensed & \begin{tabular}{@{}c@{}}$1.89 \times 10^{-22}$\\$\,^{+9.98 \times 10^{-25}}_{-9.91 \times 10^{-25}}$\end{tabular} 
  & \begin{tabular}{@{}c@{}}$0.00984$\\$\,^{+2.22 \times 10^{-11}}_{-2.20 \times 10^{-11}}$\end{tabular}
  & \begin{tabular}{@{}c@{}}$8.31 \times 10^{-15}$\\$\,^{+3.36 \times 10^{-19}}_{-3.43 \times 10^{-19}}$\end{tabular} 
  & \begin{tabular}{@{}c@{}}$2.30$\\$\,^{+5.64 \times 10^{-3}}_{-5.72 \times 10^{-3}}$\end{tabular} 
  & \begin{tabular}{@{}c@{}}$4.15$\\$\,^{+5.14 \times 10^{-02}}_{-3.10}$\end{tabular} 
  & \begin{tabular}{@{}c@{}}$4.51$\\$\,^{+7.38 \times 10^{-5}}_{-7.17 \times 10^{-5}}$\end{tabular} 
  & \begin{tabular}{@{}c@{}}$-0.301$\\$\,^{+1.66 \times 10^{-4}}_{-1.69 \times 10^{-4}}$\end{tabular} 
  & \begin{tabular}{@{}c@{}}$2.93$\\$\,^{+0.0250}_{-1.55}$\end{tabular} & -- & -- & -- & -- \\
\hline
\multirow{2}{*}{\makecell[c]{%
3: $M_\mathrm{L}=10^6$,\\
$v_{\rm eff}=50$, $L=R_\mathrm{E}$}} & Lensed & \begin{tabular}{@{}c@{}}$1.39 \times 10^{-22}$\\$\,^{+2.42 \times 10^{-23}}_{-6.86 \times 10^{-24}}$\end{tabular} 
& \begin{tabular}{@{}c@{}}$0.00984$\\$\,^{+2.27 \times 10^{-11}}_{-2.32 \times 10^{-11}}$\end{tabular} 
& \begin{tabular}{@{}c@{}}$8.31 \times 10^{-15}$\\$\,^{+3.38 \times 10^{-19}}_{-3.45 \times 10^{-19}}$\end{tabular}
&\begin{tabular}{@{}c@{}}$2.30$\\$\,^{+5.85 \times 10^{-3}}_{-5.88 \times 10^{-3}}$\end{tabular} 
& \begin{tabular}{@{}c@{}}$3.79$\\$\,^{+0.43}_{-2.94}$\end{tabular} & \begin{tabular}{@{}c@{}}$4.51$\\$\,^{+7.61 \times 10^{-5}}_{-7.68 \times 10^{-5}}$\end{tabular} 
& \begin{tabular}{@{}c@{}}$-0.301$\\$\,^{+1.69 \times 10^{-4}}_{-1.80 \times 10^{-4}}$\end{tabular}
& \begin{tabular}{@{}c@{}}$2.93$\\$\,^{+0.0235}_{-1.56}$\end{tabular} & \begin{tabular}{@{}c@{}}$716000$\\$\,^{+4.51 \times 10^{5}}_{-4.14 \times 10^{5}}$\end{tabular} 
& \begin{tabular}{@{}c@{}}$43.1$\\$\,^{+29.1}_{-21.8}$\end{tabular} & 32 & 791 \\
  & Unlensed & \begin{tabular}{@{}c@{}}$1.87 \times 10^{-22}$\\$\,^{+9.57 \times 10^{-25}}_{-9.42 \times 10^{-25}}$\end{tabular} 
  & \begin{tabular}{@{}c@{}}$0.00984$\\$\,^{+2.26 \times 10^{-11}}_{-2.21 \times 10^{-11}}$\end{tabular} 
  & \begin{tabular}{@{}c@{}}$8.31 \times 10^{-15}$\\$\,^{+3.39 \times 10^{-19}}_{-3.50 \times 10^{-19}}$\end{tabular} 
  & \begin{tabular}{@{}c@{}}$2.29$\\$\,^{+5.37 \times 10^{-3}}_{-5.58 \times 10^{-3}}$\end{tabular} 
  & \begin{tabular}{@{}c@{}}$1.09$\\$\,^{+3.11}_{-0.0467}$\end{tabular} 
  & \begin{tabular}{@{}c@{}}$4.51$\\$\,^{+7.51 \times 10^{-5}}_{-7.45 \times 10^{-5}}$\end{tabular} 
  & \begin{tabular}{@{}c@{}}$-0.301$\\$\,^{+1.74 \times 10^{-4}}_{-1.74 \times 10^{-4}}$\end{tabular} 
  & \begin{tabular}{@{}c@{}}$1.40$\\$\,^{+1.55}_{-0.0234}$\end{tabular} & -- & -- & -- & -- \\
\hline
\multirow{2}{*}{\makecell[c]{%
4: $M_\mathrm{L}=10^6$,\\
$v_{\rm eff}=500$, $L=R_\mathrm{E}$}} & Lensed & \begin{tabular}{@{}c@{}}$1.43 \times 10^{-22}$\\$\,^{+4.12 \times 10^{-23}}_{-1.10 \times 10^{-23}}$\end{tabular} 
& \begin{tabular}{@{}c@{}}$0.00984$\\$\,^{+5.70 \times 10^{-11}}_{-7.39 \times 10^{-11}}$\end{tabular} 
& \begin{tabular}{@{}c@{}}$8.31 \times 10^{-15}$\\$\,^{+3.38 \times 10^{-19}}_{-3.32 \times 10^{-19}}$\end{tabular} 
& \begin{tabular}{@{}c@{}}$2.31$\\$\,^{+5.94 \times 10^{-3}}_{-5.83 \times 10^{-3}}$\end{tabular} 
& \begin{tabular}{@{}c@{}}$3.77$\\$\,^{+0.184}_{-3.08}$\end{tabular} 
& \begin{tabular}{@{}c@{}}$4.51$\\$\,^{+7.29 \times 10^{-5}}_{-7.53 \times 10^{-5}}$\end{tabular} 
& \begin{tabular}{@{}c@{}}$-0.301$\\$\,^{+1.68 \times 10^{-4}}_{-1.66 \times 10^{-4}}$\end{tabular} 
& \begin{tabular}{@{}c@{}}$2.93$\\$\,^{+0.0202}_{-1.56}$\end{tabular} & \begin{tabular}{@{}c@{}}$1070000$\\$\,^{+3.31 \times 10^{5}}_{-2.30 \times 10^{5}}$\end{tabular} 
& \begin{tabular}{@{}c@{}}$429$\\$\,^{+270}_{-190}$\end{tabular} & 226 & 826 \\
  & Unlensed & \begin{tabular}{@{}c@{}}$1.91 \times 10^{-22}$\\$\,^{+9.90 \times 10^{-25}}_{-1.06 \times 10^{-24}}$\end{tabular} 
  & \begin{tabular}{@{}c@{}}$0.00984$\\$\,^{+2.19 \times 10^{-11}}_{-2.21 \times 10^{-11}}$\end{tabular} 
  & \begin{tabular}{@{}c@{}}$8.31 \times 10^{-15}$\\$\,^{+3.32 \times 10^{-19}}_{-3.35 \times 10^{-19}}$\end{tabular} 
  & \begin{tabular}{@{}c@{}}$2.31$\\$\,^{+5.95 \times 10^{-3}}_{-5.75 \times 10^{-3}}$\end{tabular} 
  & \begin{tabular}{@{}c@{}}$1.08$\\$\,^{+3.11}_{-0.0475}$\end{tabular} 
  & \begin{tabular}{@{}c@{}}$4.51$\\$\,^{+7.21 \times 10^{-5}}_{-7.35 \times 10^{-5}}$\end{tabular} 
  & \begin{tabular}{@{}c@{}}$-0.301$\\$\,^{+1.65 \times 10^{-4}}_{-1.71 \times 10^{-4}}$\end{tabular} 
  & \begin{tabular}{@{}c@{}}$1.39$\\$\,^{+1.56}_{-0.0234}$\end{tabular} & -- & -- & -- & -- \\
\hline
\multirow{2}{*}{\makecell[c]{%
5: $M_\mathrm{L}=10^6$,\\
$v_{\rm eff}=220$, $L=2R_\mathrm{E}$}} & Lensed & \begin{tabular}{@{}c@{}}$1.31 \times 10^{-22}$\\$\,^{+1.56 \times 10^{-23}}_{-1.37 \times 10^{-23}}$\end{tabular} 
& \begin{tabular}{@{}c@{}}$0.00984$\\$\,^{+4.32 \times 10^{-11}}_{-4.25 \times 10^{-11}}$\end{tabular} 
& \begin{tabular}{@{}c@{}}$8.31 \times 10^{-15}$\\$\,^{+5.00 \times 10^{-19}}_{-5.30 \times 10^{-19}}$\end{tabular} 
& \begin{tabular}{@{}c@{}}$2.30$\\$\,^{+8.31 \times 10^{-03}}_{-7.95 \times 10^{-03}}$\end{tabular}
& \begin{tabular}{@{}c@{}}$3.81$\\$\,^{+0.128}_{-3.05}$\end{tabular}
& \begin{tabular}{@{}c@{}}$4.51$\\$\,^{+1.10 \times 10^{-04}}_{-1.07 \times 10^{-04}}$\end{tabular} 
& \begin{tabular}{@{}c@{}}$-0.301$\\$\,^{+2.49 \times 10^{-04}}_{-2.57 \times 10^{-04}}$\end{tabular} 
& \begin{tabular}{@{}c@{}}$2.91$\\$\,^{+0.0369}_{-1.54}$\end{tabular}
& \begin{tabular}{@{}c@{}}$1060000$\\$\,^{+1.33 \times 10^{05}}_{-1.32 \times 10^{05}}$\end{tabular} 
&\begin{tabular}{@{}c@{}}$243$\\$\,^{+74.7}_{-61.0}$\end{tabular}  & 9 & 548 \\
  & Unlensed & \begin{tabular}{@{}c@{}}$1.28 \times 10^{-22}$\\$\,^{+9.78 \times 10^{-25}}_{-1.03 \times 10^{-24}}$\end{tabular} & \begin{tabular}{@{}c@{}}$0.00984$\\$\,^{+3.26 \times 10^{-11}}_{-3.13 \times 10^{-11}}$\end{tabular} & \begin{tabular}{@{}c@{}}$8.31 \times 10^{-15}$\\$\,^{+4.89 \times 10^{-19}}_{-4.91 \times 10^{-19}}$\end{tabular} & \begin{tabular}{@{}c@{}}$2.30$\\$\,^{+8.90 \times 10^{-3}}_{-8.10 \times 10^{-3}}$\end{tabular} & \begin{tabular}{@{}c@{}}$3.73$\\$\,^{+0.0690}_{-3.10}$\end{tabular} & \begin{tabular}{@{}c@{}}$4.51 \times 10^{0}$\\$\,^{+1.06 \times 10^{-4}}_{-1.17 \times 10^{-4}}$\end{tabular} & \begin{tabular}{@{}c@{}}$-0.301$\\$\,^{+2.53 \times 10^{-4}}_{-2.56 \times 10^{-4}}$\end{tabular} & \begin{tabular}{@{}c@{}}$2.92$\\$\,^{+0.0340}_{-1.55}$\end{tabular} & -- & -- & -- & -- \\
\hline
\multirow{2}{*}{\makecell[c]{%
6: $M_\mathrm{L}=10^6$,\\
$v_{\rm eff}=220$, $L=3R_\mathrm{E}$}} & Lensed & \begin{tabular}{@{}c@{}}$1.34 \times 10^{-22}$\\$\,^{+1.50 \times 10^{-23}}_{-5.58 \times 10^{-24}}$\end{tabular} & \begin{tabular}{@{}c@{}}$0.00984$\\$\,^{+3.46 \times 10^{-11}}_{-3.67 \times 10^{-11}}$\end{tabular} & \begin{tabular}{@{}c@{}}$8.31 \times 10^{-15}$\\$\,^{+4.50 \times 10^{-19}}_{-4.64 \times 10^{-19}}$\end{tabular} & \begin{tabular}{@{}c@{}}$2.30$\\$\,^{+7.56 \times 10^{-3}}_{-7.52 \times 10^{-3}}$\end{tabular} & \begin{tabular}{@{}c@{}}$0.905$\\$\,^{+3.07}_{-0.0997}$\end{tabular} & \begin{tabular}{@{}c@{}}$4.51$\\$\,^{+1.02 \times 10^{-4}}_{-9.86 \times 10^{-5}}$\end{tabular} & \begin{tabular}{@{}c@{}}$-0.301$\\$\,^{+2.27 \times 10^{-4}}_{-2.26 \times 10^{-4}}$\end{tabular} & \begin{tabular}{@{}c@{}}$1.40$\\$\,^{+1.55}_{-0.0279}$\end{tabular} & \begin{tabular}{@{}c@{}}$1050000$\\$\,^{+5.13 \times 10^{05}}_{-1.77 \times 10^{05}}$\end{tabular} & \begin{tabular}{@{}c@{}}$192$\\$\,^{+115}_{-85.1}$\end{tabular} & 1 & 593 \\
  & Unlensed & \begin{tabular}{@{}c@{}}$1.39 \times 10^{-22}$\\$\,^{+9.67 \times 10^{-25}}_{-9.68 \times 10^{-25}}$\end{tabular} & \begin{tabular}{@{}c@{}}$0.00984$\\$\,^{+2.88 \times 10^{-11}}_{-3.00 \times 10^{-11}}$\end{tabular} & \begin{tabular}{@{}c@{}}$8.31 \times 10^{-15}$\\$\,^{+4.54 \times 10^{-19}}_{-4.44 \times 10^{-19}}$\end{tabular} & \begin{tabular}{@{}c@{}}$2.30$\\$\,^{+7.62 \times 10^{-3}}_{-7.25 \times 10^{-3}}$\end{tabular} & \begin{tabular}{@{}c@{}}$0.924$\\$\,^{+3.11}_{-0.0529}$\end{tabular} & \begin{tabular}{@{}c@{}}$4.51$\\$\,^{+9.94 \times 10^{-5}}_{-1.01 \times 10^{-4}}$\end{tabular} & \begin{tabular}{@{}c@{}}$-0.301$\\$\,^{+2.26 \times 10^{-4}}_{-2.30 \times 10^{-4}}$\end{tabular} & \begin{tabular}{@{}c@{}}$1.40$\\$\,^{+1.55}_{-0.0257}$\end{tabular} & -- & -- & -- & -- \\
\hline
\end{tabular}
\end{sidewaystable*}

\begin{centering}
    \subsection{Parameter Estimation and Model Selection}
\end{centering}

To evaluate the effects of $M_\mathrm{L}$, $v_\mathrm{eff}$, and $L$, we systematically applied the control variable method in an analysis of six lensed GW signals.
Next, we perform parameter estimation for the lensed GW signal under both lensed and unlensed models using the \textsc{Bilby} \cite{2019ApJS..241...27A} framework with the \textsc{nessai} nested sampling algorithm \cite{2021PhRvD.103j3006W,2023MLS&T...4c5011W}. We then use the Bayes factor to evaluate the model distinction under different lensing parameters settings ($M_\mathrm{L}$, $v_\mathrm{eff}$, $L_0$).

In this work, for lens mass $M_\mathrm{L}=10^6$\,M$_\odot$, the prior ranges for five key parameters $(\mathcal{A}, f_0, \dot f_0, M_\mathrm{L},v_\mathrm{eff})$, are constrained by uncertainties derived from the Fisher Information Matrix (FIM) \cite{1992PhRvD..46.5236F,2008PhRvD..77d2001V}. However, the remaining parameters $(\iota, \varphi_0, \lambda, \beta, \psi)$ are set without using FIM to account for potential multimodality in their posterior distributions. In contrast, for the $10^5$\,M$_\odot$ case, the FIM produces excessively broad uncertainties for $M_\mathrm{L}$ and $v_\mathrm{eff}$. We therefore adopt physical priors for the two parameters instead. Specifically, we assign a logarithmic prior for $M_\mathrm{L}\in[10, 10^6]$\,M$_\odot$ and a uniform prior for $v_\mathrm{eff}\in[0, 600]$\,km/s. \Cref{tab:1} summarizes the parameter estimations for six lensed GW cases under various lens configurations.
\begin{figure*}[tp]
    \centering
    \includegraphics[width=\linewidth]{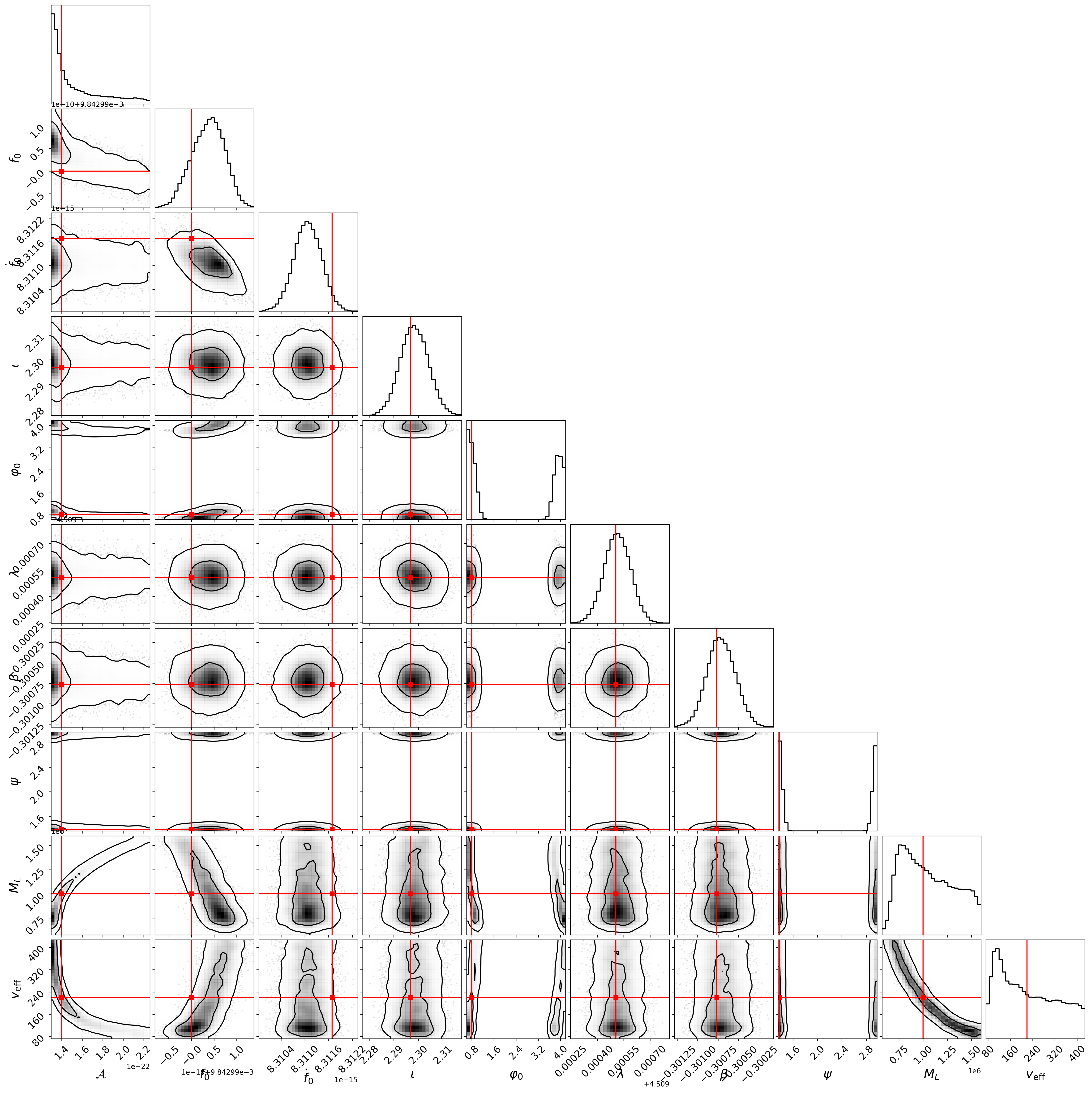}
    \caption{Parameter estimation results for the lens model in case 2. In this case, $M_\mathrm{L}=10^6$\,M$_\odot$, $L=R_\mathrm{E}$, $v_\mathrm{eff}=220$\,km/s}
    \label{fig:6}
\end{figure*}

In case 1 and case 3, all parameters except $M_\mathrm{L}$ are fixed to examine the impact of variations in $M_\mathrm{L}$ on the inferred parameters and the model distinction. The results show that, given the current prior settings, the lensed and unlensed models remain statistically indistinguishable when $M_\mathrm{L}$ is $10^5$\,M$_\odot$. In cases 2, 5 and 6, we vary $L$ from $R_\mathrm{E}$ to $3R_\mathrm{E}$ while keeping other parameters constant. The SNR is lower in Case 5 than in Case 6, despite its smaller $L$,  because Case 5 has a smaller magnification factor $|F(t)|$. The results demonstrate that when $L=3R_\mathrm{E}$, the lensed and unlensed models become statistically indistinguishable. In cases 2, 3, and 4, we varied $v_\mathrm{eff}$ while keeping all other parameters constant. The results indicate that within the range of $v_\mathrm{eff}$ that we considered, the lensed and unlensed models can be distinguished. This confirms the result of \Cref{fig:4}, which shows that the effect of $v_\mathrm{eff}$ on the waveform is weak.
\begin{figure*}[tp]
    \centering
    \includegraphics[width=\linewidth]{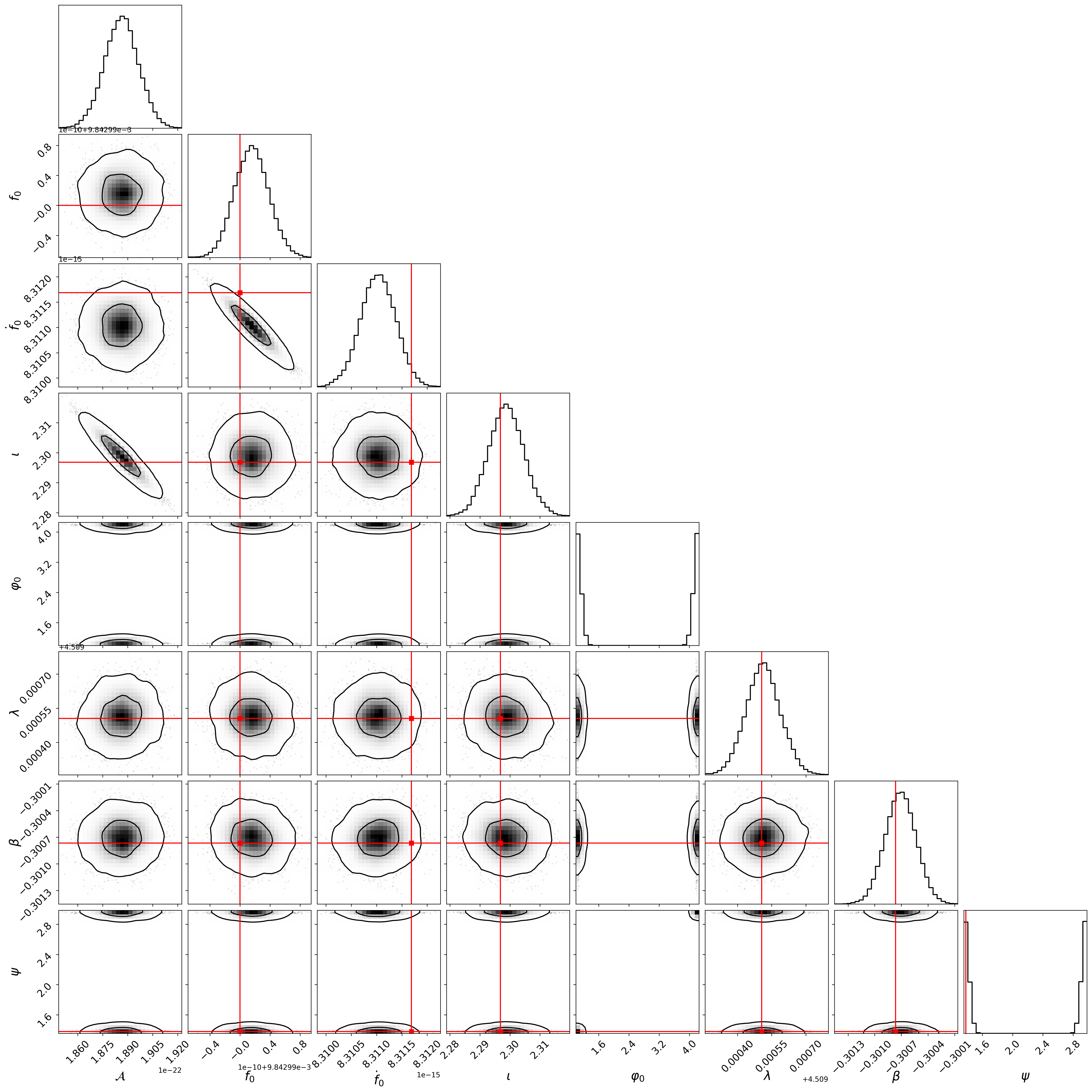}
    \caption{Parameter estimation results for the unlens model in case 2.}
    \label{fig:7}
\end{figure*}

In \Cref{fig:6} and \Cref{fig:7} we show the posterior distributions obtained under the lensed and unlensed models for Case 2. Consistent with the analysis above, the posterior results indicate that $\mathcal{A}$, $M_\mathrm{L}$, and $v_\mathrm{eff}$ are poorly constrained. In addition, the posterior distributions reveal a clear tripartite correlation among $M_\mathrm{L}$, $\mathcal{A}$, and $v_\mathrm{eff}$. This complex degeneracy arises because variations in both $M_\mathrm{L}$ and $v_\mathrm{eff}$ can jointly influence the amplification factor, which in turn combines with the GW amplitude $\mathcal{A}$ to form the total amplitude. Within the wave optics regime, these dependencies make it difficult to effectively limit $M_\mathrm{L}$ and $v_\mathrm{eff}$. Furthermore, when lensing effects are taken into account, the GW amplitude $\mathcal{A}$ is effectively magnified. Consequently, as shown in \Cref{fig:7}, the GW amplitude $\mathcal{A}$ inferred with the unlensed model is larger than that obtained with the lensed model and the truth value. This bias arises because, in the absence of lensing, the model compensates for the missing magnification by favoring a higher GW amplitude $\mathcal{A}$.  
In addition, we find that in \Cref{fig:7} the $\varphi_0$ inferred under the unlensed model exhibits a pronounced distortion. According to Appendix~\ref{appA}, the observed phase distortion arises because the total phase $\varPhi_\mathrm{L}(t)$ is a combination of the GW phase $\varphi(t)$ and the phase term induced by magnification $\theta_\mathrm{L}(t)$. This behavior further illustrates the degeneracy between $M_\mathrm{L}$, $v_\mathrm{eff}$ and $\mathcal{A}$.

All of the above, our analysis and results show that the effect of $v_\mathrm{eff}$ on distinguishing lensing times is weak. At the same time, $M_\mathrm{L}$ and $L$ play a dominant role in distinguishing lensing events. Meanwhile, there is a degeneracy among $\mathcal{A}$, $M_\mathrm{L}$, and $v_\mathrm{eff}$. In addition, lensing effects lead to distorted estimates in the estimates of $\mathcal{A}$ and $\varphi_0$ under the unlensed model. The findings highlight the importance of considering lensing effects when analyzing GW signals from DWD systems in the Taiji frequency band.

\begin{centering}
    \section{Conclusion}
    \label{Sec:Conclusion}
\end{centering}

In this study, we investigated the effects of GW lensing on DWD systems in the Taiji frequency band under the PML model, focusing on wave-optics signatures induced by lensing objects such as black holes, globular clusters, and dark matter subhalos, while explicitly accounting for the relative motion between the source and the lens. We simulate the GW data within the fast-slow decomposition approach and the Taiji orbital data over a four year period. We used the controlled variable approach to analyze the impacts of $M_\mathrm{L}$, $v_\mathrm{eff}$, and $L$ on the GW signal, which allows us to characterize and differentiate different lens signatures within the Bayesian framework.

In this work, we analyzed six representative cases and the results of the posterior distributions are summarized in \Cref{tab:1}. The results highlight the significant role that $M_\mathrm{L}$ and $L$ play in the identification of lensing events. By contrast, the impact of $v_\mathrm{eff}$ on the lensed signal is found to be negligible. Specifically, our sensitivity analysis indicates that lensing events with $M_\mathrm{L}=10^5$\,M$_\odot$ are statistically indistinguishable from unlensed signals, while they become clearly identifiable at $10^6$\, M$_\odot$. Furthermore, for $M_\mathrm{L}=10^6$\,M$_\odot$, the lens signatures become indistinguishable from the unlensed model at $L=3R_\mathrm{E}$. With respect to $v_\mathrm{eff}$, it has a negligible impact on the distinguishability of lensing events within the range of parameters considered. Our results suggest that for $M_\mathrm{L}$ exceeding $10^5$\,M$_\odot$, globular clusters, supermassive black holes, and dark matter subhalos are the most plausible candidates. Additionally, the posterior results show that $M_\mathrm{L}$, $\mathcal{A}$, and $v_\mathrm{eff}$ exhibit a strong correlation, indicating the difficulty of independently constraining these parameters due to the degeneracy introduced by lens effects. 

Finally, it should be noted that our study makes some idealized assumptions. One of these is considering a static single lens object, while in reality, there may be multiple lensing bodies, and they could be dynamic. Additionally, treating globular clusters and dark matter subhalos as PML is also an idealization. In the future, joint electromagnetic observations could help resolve the degeneracy among $M_\mathrm{L}$, $\mathcal{A}$, and $v_\mathrm{eff}$, thus improving the inference of parameters.\\

\begin{center}
\textbf{ACKNOWLEDGMENTS}
\end{center}
\noindent
\indent This work is supported by the National Key Research and Development Program of China (Grant No.~2025YFE0217300), and the International Partnership Program of the Chinese Academy of Sciences (Grant No.~025GJHZ2023106GC).

\appendix
\begin{centering}
    \section{Deduction of the Fast TDI Response for Lensed DWD Signals}
    \label{appA}
\end{centering}

In this appendix, we derive an efficient expression for the TDI response of lensed DWD signals suitable for Bayesian analysis, following the ideas proposed in reference \cite{2025PhRvD.112j2007C,2007PhRvD..76h3006C}.

TDI response for generic GW signal, detector orbit model, TDI variable in the time domain: 
\begin{equation}\mathrm{TDI}(t)=\sum_{ij\in\mathcal{I}_2}\mathbf{P}_{ij}y_{ij}(t),\end{equation}
where $\mathcal{I}_2\equiv\{12,23,31,21,32,13\}$, $\mathbf{P}_{ij}$ represents the polynomial of time-delay operators, whose expression for specific TDI channels will be given below, $y_{ij}(t)$ denotes the (single-arm) response of the laser link spacecraft$_j\rightarrow$ spacecraft$_i$ in units of fractional frequency shift:
\begin{equation}
\begin{aligned}
y_{ij}(t) &\equiv \frac{\nu_{\mathrm{receive}} - \nu_{\mathrm{send}}}{\nu_{\mathrm{send}}} \\
&\approx \frac{1}{2\left(1 - \boldsymbol{\hat{k}} \cdot \boldsymbol{\hat{n}}_{ij}(t)\right)} \times \\
&\quad \left[ H_{ij}\left(t - d_{ij}(t) - \boldsymbol{\hat{k}} \cdot \boldsymbol{R}_j(t)\right) \right. \\
&\quad \left. - H_{ij}\left(t - \frac{\boldsymbol{\hat{k}} \cdot \boldsymbol{R}_i(t)}{c}\right) \right],
\end{aligned}
\end{equation}
where $\boldsymbol{R}_{i}(t)$ is the position of the spacecraft$_i$ in the Solar System Barycenter (SSB) frame, $d_{ij}(t)$ is the light travel time from the spacecraft$_j$ to the spacecraft$_i$, and $\boldsymbol{\hat{n}}_{ij}(t)$ represents the unit vector along this arm. The projection of the GW tensor onto the arm $ij$ reads:
\begin{equation}H_{ij}(t)\equiv h_+(t)\zeta_{+,ij}(t)+h_\times(t)\zeta_{\times,ij}(t),\end{equation}
where
\begin{equation}\begin{aligned}&\zeta_{+,ij}(t)=\cos(2\psi)\xi_{+,ij}(t)+\sin(2\psi)\xi_{\times,ij}(t),\\&\zeta_{\times,ij}(t)=-\sin(2\psi)\xi_{+,ij}(t)+\cos(2\psi)\xi_{\times,ij}(t),\\&\xi_{+,ij}(t)=[\boldsymbol{\hat{n}}_{ij}(t)\cdot\boldsymbol{\hat{u}}]^2-[\boldsymbol{\hat{n}}_{ij}(t)\cdot\boldsymbol{\hat{v}}]^2,\\&\xi_{\times,ij}(t)=2\left[\boldsymbol{\hat{n}}_{ij}(t)\cdot\boldsymbol{\hat{u}}\right]\left[\boldsymbol{\hat{n}}_{ij}(t)\cdot\boldsymbol{\hat{v}}\right].\end{aligned}\end{equation}
Namely, the $\zeta_s$ and $\xi_s$ are the antenna pattern functions in the source frame and the SSB frame, respectively, and $\psi$ stands for the polarization angle. For a GW source located at Ecliptic longitude $\lambda$ and Ecliptic latitude $\beta$, the Cartesian coordinate components of the unit vectors $\boldsymbol{\hat{u}},\boldsymbol{\hat{v}},$ and $\boldsymbol{\hat{k}}$ are 
\begin{equation}
\begin{aligned}
\hat{\mathbf u} &= [\sin\lambda,-\cos\lambda,0],\\
\hat{\mathbf v} &= [-\sin\beta\cos\lambda,-\sin\beta\sin\lambda,\cos\beta],\\
\hat{\mathbf k} &= -[\cos\beta\cos\lambda,\cos\beta\sin\lambda,\sin\beta].
\end{aligned}
\end{equation}
Through the application of the operator $\mathbf{P}_{ij}$ and linear combinations, any TDI variable can be composed of observables in a single-arm. Taking the second-generation Michelson channel $X_{2}$ as an example, the corresponding $\mathbf{P}_{ij}$ operators are:
{
\begin{equation}
\begin{aligned}&\mathbf{P}_{12}=1-\mathbf{D}_{131}-\mathbf{D}_{13121}+\mathbf{D}_{1213131},\\&\mathbf{P}_{23}=0,\\&\mathbf{P}_{31}=-\mathbf{D}_{13}+\mathbf{D}_{1213}+\mathbf{D}_{121313}-\mathbf{D}_{13121213},\\&\mathbf{P}_{21}=\mathbf{D}_{12}-\mathbf{D}_{1312}-\mathbf{D}_{131212}+\mathbf{D}_{12131312},\\&\mathbf{P}_{32}=0,\\&\mathbf{P}_{13}=-1+\mathbf{D}_{121}+\mathbf{D}_{12131}-\mathbf{D}_{1312121}.\end{aligned}\end{equation}
}
where the delay operator $\mathbf{D}_{i_1i_2}$ acting on any time function $f(t)$ results in $\mathbf{D}_{i_1i_2}f(t)\equiv f[t-d_{i_1i_2}(t)]$. The multiple delay operator is defined as $\mathbf{D}_{i_1i_2i_3...}f(t)\equiv\mathbf{D}_{i_1i_2}\mathbf{D}_{i_2i_3}...f(t)$. The definitions for $Y$, $Z$ channels can be obtained by cyclically permuting the indices: $1\to2,2\to3,3\to1$. For Bayesian analysis using multiple TDI channels, it is more convenient to use the quasi-noise-orthogonal combinations $A_\mathrm{2}$, $E_\mathrm{2}$, and $T_\mathrm{2}$:
\begin{equation}
\begin{aligned} & A_2\equiv\frac{Z_2-X_2}{\sqrt{2}},\\  & E_2\equiv\frac{X_2-2Y_2+Z_2}{\sqrt{6}},\\&T_2\equiv\frac{X_2+Y_2+Z_2}{\sqrt{3}}.
\end{aligned}
\end{equation}

The GW signal from a DWD is given by \Crefrange{Eq15}{Eq18}, and the TDI response for a DWD signal reads:
\begin{equation}
\begin{aligned}
    \mathrm{TDI}(t)=&\sum_{ij}\frac{1}{2(1-\hat{\boldsymbol{k}}\cdot\hat{\boldsymbol{n}}_{ij})}\sum_P\zeta_{P,ij}\mathbf{P}_{ij}\\&\times\left[h_P\left(t-d_{ij}-\frac{\hat{\boldsymbol{k}}\cdot\boldsymbol{R}_j}{c}\right)-\right.\\&\left. h_P\left(t-\frac{\hat{\boldsymbol{k}}\cdot\boldsymbol{R}_i}{c}\right)\right],
    \label{Eq.A8}
\end{aligned}
\end{equation}
we further decompose $\mathbf{P}_{ij}$ as:
\begin{equation}
\mathbf{P}_{ij}=\sum_{I_{ij}}K_{I_{ij}}\mathbf{D}_{I_{ij}},\quad\mathbf{D}_{I_{ij}}f(t)=f\begin{pmatrix}t-d_{I_{ij}}\end{pmatrix}.
\label{Eq.A9}
\end{equation}
where $K_{I_{ij}}$ can be either $+1$ or $-1$ for the Michelson $X_2$ TDI channel.
Then, we consider the lens effect, and then \Cref{Eq.A8} becomes:
\begin{equation}
\begin{aligned}
    \mathrm{TDI}(t)=&\sum_{ij}\frac{1}{2(1-\hat{\boldsymbol{k}}\cdot\hat{\boldsymbol{n}}_{ij})}\sum_P\zeta_{P,ij}\mathbf{P}_{ij}\\&\times\left[h^L_P\left(t-d_{ij}-\frac{\hat{\boldsymbol{k}}\cdot\boldsymbol{R}_j}{c}\right)-\right.\\&\left. h^L_P\left(t-\frac{\hat{\boldsymbol{k}}\cdot\boldsymbol{R}_i}{c}\right)\right],
\end{aligned}
\end{equation}
where $h_P^L(t)=F(f_0,t)h_P(t)$. 
For any delay operation $\mathbf{D}$ in $\mathbf{P}$, the delay times are of order $<10^{2}s$, one has $\mathbf{D}\boldsymbol{\hat{n}}_{ij}(t)\approx\boldsymbol{\hat{n}}_{ij}$. So, the calculation of the TDI response reduces to the calculation of $\mathbf{F}(t)\mathbf{D}h^L_{P}(t)$, where $\mathbf{F}(t)$ is the abstract form of the pattern function, and $\mathbf{D}$ is induced by $\mathbf{P}$ or Doppler terms (the terms $\boldsymbol{\hat{k}}\cdot\boldsymbol{R}/c$). Then, we apply \Cref{Eq.A9} and obtain:
{
\begin{equation}
\begin{aligned}\mathrm{TDI}(t)&=\Re\left\{\sum_{ij}\sum_{I_{ij}}\frac{K_{I_{ij}}}{2(1-\hat{\boldsymbol{k}}\cdot\hat{\boldsymbol{n}}_{ij})}\left(\sum_{P}A_{P}\zeta_{P,ij}\right)\right.\\&\times\left.\left[\left|F\left(t-d_{I_{ij}}^{\mathrm{send}}\right)\right|e^{i\left[\varPhi_\mathrm{L}(t-d_{I_{ij}}^{\mathrm{send}})-\varPhi_\mathrm{L}(t)\right]}\right.\right.\\&-\left.\left.\left|F\left(t-d_{I_{ij}}^{\mathrm{recv}}\right)\right|e^{i\left[\varPhi_\mathrm{L}(t-d_{I_{ij}}^{\mathrm{recv}})-\varPhi_\mathrm{L}(t)\right]}\right]e^{i\varPhi_\mathrm{L}(t)}\right\}
\label{Eq.A11}
\end{aligned}\end{equation}
}
where we define $d_{I_{ij}}^{\mathrm{send}}=d_{I_{ij}}+d_{ij}+\hat{\boldsymbol{k}}\cdot\boldsymbol{R}_{j}/c,\,\,d_{I_{ij}}^{\mathrm{recv}}=d_{I_{ij}}+\hat{\boldsymbol{k}}\cdot\boldsymbol{R}_{i}/c$, $\left|F(t)\right|$ is the amplitude of the amplification factor, $\varPhi_\mathrm{L}(t) = \varphi(t) + \theta_\mathrm{L}(t)$ and $\theta_\mathrm{L}(t)$ is the phase of the amplification factor. We employ the fast-slow decomposition approach \cite{2007PhRvD..76h3006C,2025PhRvD.112j2007C} to calculate \Cref{Eq.A11}. Notably, the terms $\mathbf{F}(t)$, $\sum_PA_P\zeta_{P,ij}$, $F(t-d)$, $e^{i[\varPhi_\mathrm{L}(t-d)-\varPhi_\mathrm{L}(t)]}$, and $\theta_\mathrm{L}(t)$ vary slowly with time and are collectively referred to as the "slow part".
The "fast part" is governed by the GW phase $\varphi(t)$.
Then, since the GW from DWD occupies only a narrow frequency band around $f_\mathrm{c} = f_0$, and considering the lensed effect, \Cref{Eq.A11} can be rewritten as an abstract form for the resulting time-domain TDI response:
\begin{equation}
\mathrm{TDI}(t)=\Re\left\{A_\mathrm{L}^{\mathrm{TDI}}(t)e^{i\Delta\Phi_\mathrm{L}(t)}e^{i\Phi_\mathrm{c}(t)}\right\},
\label{Eq.A12}
\end{equation}
where $\Phi_\mathrm{c}(t)\equiv2\pi f_\mathrm{c}t,\quad\Delta\Phi_\mathrm{L}(t)\equiv\Phi_\mathrm{L}^{\mathrm{TDI}}(t)+\varPhi_\mathrm{L}(t)-\Phi_\mathrm{c}(t)$, $\Phi_\mathrm{L}^{\mathrm{TDI}}(t)\equiv e^{i[\varPhi_\mathrm{L}(t-d)-\varPhi_\mathrm{L}(t)]}$.
As explained above, the terms $A_\mathrm{L}^{\mathrm{TDI}}(t)$ and $e^{i\Delta\Phi_\mathrm{L}(t)}$ vary slowly over time. They are therefore referred to as the "slow part" and can be represented by a relatively sparse time grid. On the other hand, the term $e^{i\Phi_\mathrm{c}(t)}$ varies rapidly with time and is considered the "fast part". However, for a single harmonic (such as a DWD signal), its Fourier transform can be calculated analytically and reduces to a Dirac -$\delta$ function. 
In summary, our strategy for calculating the TDI response is as follows: first, we select a sparse grid (e.g., 256 points per year). Next, we compute the arm vectors $\boldsymbol{\hat{n}}_{ij}$, spacecraft positions $\boldsymbol{R}_i$, delay times $d_{I_{ij}}^{\mathrm{send}}$ and $d_{I_{ij}}^{\mathrm{recv}}$ on the grid. Then, additional terms related to the given DWD source parameters are calculated on the grid. The delay terms and laser arms are subsequently summed to obtain the amplitude $A_\mathrm{L}^{\mathrm{TDI}}(t)$ and $\Phi_\mathrm{L}^{\mathrm{TDI}}(t)$. Finally, linear interpolation is applied to obtain the full set of values for $A_\mathrm{L}^{\mathrm{TDI}}(t)$, $\Phi_\mathrm{L}^{\mathrm{TDI}}$ and $\varPhi_\mathrm{L}(t)$ across the entire time grid, which are then combined according to the abstract formula to generate the final TDI response.

Since Bayesian statistics are defined in the frequency domain, it would be more convenient to use the Fourier transform of \Cref{Eq.A12}.
Taking the Fourier transform of the above expression and applying the convolution theorem, we get:
\begin{multline}
\widetilde{\mathrm{TDI}}(f) = \frac{1}{2}\mathcal{F}\left[A_\mathrm{L}^{\mathrm{TDI}}(t)e^{i\Delta\Phi_\mathrm{L}(t)}e^{i\Phi_\mathrm{c}(t)}\right] \\
= \frac{1}{2}\mathcal{F}\left[A_\mathrm{L}^{\mathrm{TDI}}(t)e^{i\Delta\Phi_\mathrm{L}(t)}\right] * \mathcal{F}\left[e^{i\Phi_\mathrm{c}(t)}\right].
\end{multline}
For long-duration data, the Fourier transform of the monochromatic wave $e^{i\Phi_{c}}$ can be well approximated by a Dirac -$\delta$ function. Thus, 
\begin{equation}\widetilde{\mathrm{TDI}}(f)=\frac{1}{2}\mathcal{F}\left[A_\mathrm{L}^{\mathrm{TDI}}(t)e^{i\Delta\Phi_\mathrm{L}(t)}\right](f-f_\mathrm{c}).\end{equation}
Notice that the desired result is only limited to a narrow bandwidth around $f_\mathrm{c}$, thus calculating the former term on a sparse time grid (with 1024 data points, four years) and then FFT should be enough. The resulting $\widetilde{\mathrm{TDI}}(f)$ contains 1024 frequency points centered on $f_\mathrm{c}$ with a frequency resolution of $\frac{1}{T_{obs}}$.

\bibliographystyle{apsrev4-2}
\bibliography{apssamp}

\end{document}